\documentclass[]{aa} % for the letters  
\usepackage{graphicx}
\usepackage{txfonts}
\usepackage{natbib}
\usepackage{longtable}
\usepackage{lscape}
\usepackage{color}
\providecommand{\sorthelp}[1]{}

\begin{document}
\title{ Dust emission, extinction, and scattering in LDN~1642 
\thanks{ The paper is based on observations collected at the European
Southern Observatory under ESO programme 090.C-0603.  {\it Herschel}
is an ESA space observatory with science instruments provided by
European-led Principal Investigator consortia and with important
participation from NASA.} }

\author{Mika Juvela\inst{1}, 
        Sharma Neha\inst{1,2},
        Emma Mannfors\inst{1},
        Mika Saajasto\inst{1},
        Nathalie Ysard\inst{3},
        Veli-Matti Pelkonen\inst{4}
}

\institute{
Department of Physics, P.O.Box 64, FI-00014, University of Helsinki,
Finland, {\em mika.juvela@helsinki.fi}
\and
Finnish Centre for Astronomy with ESO (FINCA), FI-20014 University of
Turku, Finland
\and
%% IAS, Universit{\'e} Paris-Sud, 91405 Orsay Cedex, France
Universit{\'e} Paris-Saclay, CNRS, Institut d’Astrophysique Spatiale,
91405, Orsay, France
\and
Institut de Ci\`{e}ncies del Cosmos, Universitat de Barcelona, IEEC-UB, Mart\'{i} i Franqu\`{e}s 1, E08028
Barcelona, Spain
}

\authorrunning{First Author et al.}
\date{Received September 15, 1996; accepted March 16, 1997}

\abstract{
% Background
LDN~1642 is a rare example of a star-forming high-latitude molecular
cloud. The dust emission of LDN~1642 has already been studied
extensively in the past, but its location also makes it a good target
for studies of light scattering.
}{
% Aims
We wish to study the near-infrared (NIR) light scattering in LDN~1642,
its correlation with the cloud structure, and the ability of dust
models to simultaneously explain observations of sub-millimetre dust
emission, NIR extinction, and NIR scattering.
}{
% Methods
We use observations made with the HAWK-I instrument to measure the NIR
surface brightness and extinction in LDN~1642. These data are compared
with Herschel observations of dust emission and, with the help of
radiative transfer modelling, with the predictions calculated for
different dust models.
}{
% Results
We find for LDN~1642 an optical depth ratio $\tau(250\,\mu{\rm
m})/\tau(J)\approx 10^{-3}$, confirming earlier findings of
enhanced sub-millimetre emissivity. The relationships between the
column density derived from dust emission and the NIR colour excesses
is linear and consistent with the shape of the standard NIR extinction
curve. The extinction peaks at $A_J=2.6$\,mag, the NIR surface
brightness remaining correlated with $N({\rm H}_2)$ without saturation.
Radiative transfer models are able to fit the sub-millimetre data with
any of the tested dust models. However, these predict a NIR extinction
that is higher and a NIR surface brightness that is lower than based
on NIR observations. If the dust sub-millimetre emissivity is rescaled
to the observed value of $\tau(250\,\mu{\rm m})/\tau(J)$, dust
models with high NIR albedo can reach the observed level of NIR
surface brightness. The NIR extinction of the models tends to be
higher than in the direct extinction measurements, which also is
reflected in the shape of the NIR surface brightness spectra.
}{
% Conclusions
The combination of emission, extinction, and scattering measurements
provides strong constraints on dust models. The observations of
LDN~1642 indicate clear dust evolution, including a strong increase in
the sub-millimetre emissivity, not yet fully explained by the
current dust models.
%%Remaining discrepancies between
%%observations and models can be explained by dust property variations
%%as a function of density.
%
}


\maketitle

\section{Introduction}

Dust is central to the physics of the interstellar medium (ISM) and
important for the heating of the gas, the formation of H$_{\rm 2}$
molecules, and the shielding of more complex molecules from the
interstellar radiation field (ISRF). Dust emission is used as a tracer
of the ISRF, star formation (SF) activity, and ISM mass, all affected by
the dust properties and its abundance. It is thus essential to know
the properties of interstellar dust that affect their light scattering
and thermal dust emission. 

Coreshine and cloudshine refer to excess signal detected in the
infrared. Cloudshine is caused by the scattering of the interstellar
radiation field from the clouds at near-infrared (NIR) wavelengths
\citep{Foster2006,Ysard2016}, while coreshine is caused by scattered
photons from deeper within the dense cores, visible in the
mid-infrared \citep[MIR;][]{Steinacker2010,Pagani2010}. Coreshine and
cloudshine provide a way to study the growth of grains in the dense
interstellar medium \citep{Ysard2018}. Both are affected by changes in
dust scattering efficiency, which may be related to the surface
irregularity of grains, changing grain size or fluffiness,
coagulation, and ice coating
\citep{Ossenkopf1993,Stepnik2003,Ridderstad2010,
Ormel2011,Ysard2013A,Kohler2015,Min2016,Ysard2016}. 

Dust is heated by stellar UV-visible radiation, and the absorbed
energy is radiated away in a range from MIR to far-infrared (FIR) and
millimetre wavelengths. The observed variation of the MIR-to-FIR
ratios is believed to be due to dust grain evolution, grain growth,
and ice mantle formation with increasing density
\citep{Ormel2009,Boogert2015,Kohler2015}. These  changes are reflected in the
dust spectral energy distribution (SED), which thus provides important
clues on dust evolution processes such as grain coagulation and
fragmentation \citep{Compiegne2011}. 

Stars form from collapsing clouds of dense interstellar gas and dust, dust
emission being an important tracer of the process. Filamentary structures are
common in molecular clouds (MCs)
\citep[e.g.][]{Menshchikov2010,Andre2010,Arzoumanian2011,
Hennemann2012,Juvela2012_filaments,
Malinen2012,Palmeirim2013,Andre2014,Wang2015,Andre2019} and they fragment to
subparsec-scale cores, which may subsequently lead to the formation of young
stellar objects \citep[YSOs;][]{Kirk2013,Offner2014,Konyves2015}. SF is studied
with both dust and molecular line observations and the latter are essential for
investigations of cloud stability, kinematics, and chemistry
\citep{Motte1998,Bergin2007,Enoch2007,Pattle2017}. However, dust is a central
tool also in the study of SF processes. The far-IR dust emission, often
approximated as modified blackbody (MBB) emission, traces not only the
column density but through the dust temperature also the ISRF changes that
are associated with the general SF activity \citep{Sadavoy2013,Planck2013}. At
shorter wavelengths, under 100\,$\mu$m, emission from hot dust is important for
the detection and characterisation of young stellar objects (YSOs)
\citep{Lada1987,Benedettini2018}. 

High-latitude clouds ($|b| > 30$\degr) are a fairly rare class of
interstellar clouds \citep{Dutra2002,McGehee2008}. They are typically
nearby objects with low column densities and no star formation. There
are only a handful of high-latitude clouds with more molecular
material and some low-mass SF activity \citep{Lynds1962,Malinen2014}.
These are excellent targets for studying SF triggered by supernovae
and stellar winds, as SF due to gravitational collapse is less likely
\citep{Elmegreen1998,McGehee2008}. Because of the low levels of
line-of-sight (LOS) confusion, they are good targets also for studies
into the dust properties in interstellar clouds.

\begin{table}
\caption{Properties of L1642. \label{tbl:L1642_properties}}
\centering
\begin{tabular}{llllll}
\hline\hline
$\ell$ & \textit{b} & $\alpha_{\rm 2000}$ & $\delta_{\rm 2000}$ & Distance & $A_{\rm v}$\\
(\degr) & (\degr) & (\degr) & (\degr) & (pc) & (mag) \\
\hline
210.9\tablefootmark{a} & -36.55\tablefootmark{a} & 68.75\tablefootmark{a} & -14.25\tablefootmark{a} & 140\tablefootmark{b} & 2.0\tablefootmark{c}\\
\hline
\end{tabular}
\tablefoot{ 
\tablefoottext{a}{From \citet{Malinen2014}.} 
\tablefoottext{b}{From \citet{Kuntz1997,Sfeir1999}.}
\tablefoottext{c}{From \citet{McGehee2008}.
The $A_V$ value is estimated from CO, H{\sc I}, and FIR surveys using $E$($B$-$V$)
values of \citet{Dutra2002}.
}}
\end{table}

LDN~1642, also referred to as MBM 20 and G210.90-36.55
\citep{Lynds1962,Magnani1985,GCC-III}, is one of the star-forming
high-latitude clouds (Table \ref{tbl:L1642_properties}). It is
gravitationally bound and hosts three YSO systems \citep{McGehee2008,
Malinen2014}.  LDN~1642 is part of a larger ($> 4$\degr) cometary HI
cloud, with an over 5\degr\ long tail toward the Galactic plane
\citep{Gir1994,Alcala2008}. The cloud is projected on the Orion-Eridanus
bubble \citep{Brown1995}, with which it may be interacting
\citep{Lehtinen2004}. It is also located $\sim$10\degr\, from the
reflection nebula IC 2118 (the Witch Head nebula)
\citep{Kun2001,Alcala2008}. The cloud structure and the large-scale
magnetic field of LDN~1642 are linked, the magnetic field possibly
affecting the cloud evolution \citep{Malinen2016}. There is a clear
change from magnetic-field-aligned to perpendicular structures around
a column density of $N_{\rm H} = 1.6\times 10^{21}$\,cm$^{-2}$
\citep{Malinen2016}. The light scattering in LDN~1642 at optical
wavelengths has already been studied in
\citet{Mattila2007,Mattila2018}.

LDN~1642 contains several denser regions, named by \citet{Lehtinen2004} as A1,
A2, B, and C. \citet{Malinen2014} divide region B into two subregions, B1 and B2,
due to an intensity maximum separate from the main clump. Three pre-main-sequence
objects are associated with LDN~1642. Two of these, IRAS 04327-1419 = L1642-1 (V*
EW Eri, HBC 413) and IRAS 04325-1419 = L1642-2 (HBC 410) were discovered by
\citet{Sandell1987}. We refer to these sources as B1 and B2, respectively,
and their locations are indicated in Fig.~\ref{fig:colden}. Both are faint
binary stars. The primary of L 1642-1 is classified as a Type II YSO T-Tauri star
of spectral class K7IV \citep{Sandell1987,Malinen2014} and the primary of L
1642-2 as a flat-spectrum YSO M0 class H$\alpha$-emission star
\citep{Liljestrom1989,Malinen2014}. A weak, bipolar outflow has been found around
B2, and a Herbig-Haro object (HH123) originates from it
\citep{Liljestrom1989,Reipurth1990}. 
The 2MASS point source 2MASS J04351455-1414468 was originally
classified as a potential foreground dwarf star \citep{Cruz2003} but
is now considered to be a Type III YSO associated with LDN~1642
\citep{Malinen2014}. We refer to this object as B3
(Fig.~\ref{fig:colden}).

In this paper, we study the cloud LDN~1642 by combining Herschel
satellite data with new NIR observations. With the help of radiative
transfer (RT) modelling, we test the ability of dust models to
consistently predict in LDN~1642 all three aspects of dust
observations: sub-millimetre emission, NIR scattering, and NIR 
extinction. Its location at a small distance and high above the
Galactic plane makes LDN~1642 a good target for this study. 

The contents of the paper are the following. The observations at
sub-millimetre, NIR, and optical wavelengths are presented in
Sect.~\ref{sect:observations} and the main observational results in
Sect.~\ref{sect:results}. Section~\ref{sect:RT} describes the RT
modelling, where results are shown for dust emission in
Sect.~\ref{sect:mod_Herschel} and for NIR scattering in
Sect.~\ref{sect:extended} and Sect.~\ref{sect:PS_models}. We discuss
the results in Sect.~\ref{sect:discussion} before presenting the final
conclusions in Sect.~\ref{sect:conclusions}.

\section{Observations} \label{sect:observations}

\subsection{Dust emission} \label{sect:Herschel_observations}

We use the pipeline-reduced Herschel observations from the Herschel science
archive\footnote{http://archives.esac.esa.int/hsa/whsa/}. Of the observations
made with the PACS instrument \citep{Poglitsch2010}, we use the 160\,$\mu$m data
(level 2.5 data products, observation ID numbers 1342225212 and 1342225213), the
maps produced with the Scanamorphos algorithm \citep{Roussel2013}. The PACS
100\,$\mu$m data show little extended emission and even the embedded sources are
associated with little extended emission. The observations with the SPIRE
instrument \citep{Griffin2010} cover the wavelengths 250\,$\mu$m, 350\,$\mu$m,
and 500\,$\mu$m (observation ID 1342216940). The data correspond to
extended-source calibration and, with a comparison to Planck data, have already
been zero-point corrected by the pipeline \citep[see][]{Bernard2010}. However, we
analyse the data using background subtraction, which makes the results
independent of the zero-point accuracy. The background values were estimated as
the average intensity within 3.7 arcmin of the position RA=4:34:49.35,
DEC=-14:26:21.70. For the background determination, the maps were also first
convolved to a common 41$\arcsec$ resolution. Here, and later in the analysis, we
use for SPIRE the convolution kernels provided by \citet{Aniano2011}.

\begin{figure*}
\includegraphics[width=18.0cm]{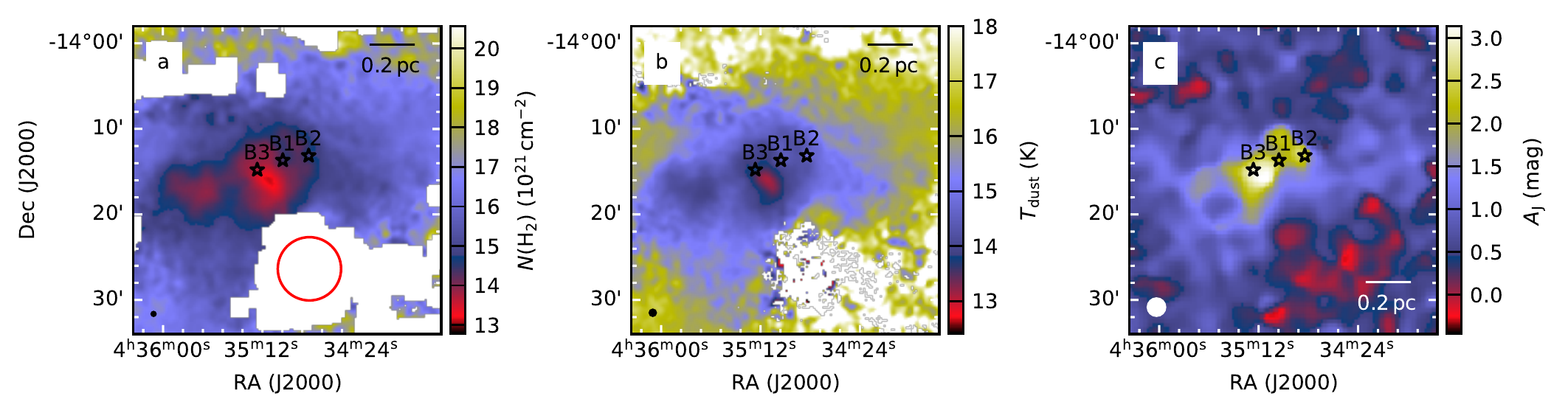}
%
%\sidecaption
\caption{
Column density $N({\rm H}_2)$  (frame a), dust temperature $T_{\rm dust}$
(frame b), and $J$-band extinction $A_J$ (frame c) of LDN\,1642. 
The resolution of the maps is indicated by the circles in the lower left
corner. 
The
values of $N({\rm H}_2)$ and $T_{\rm dust}$ are based on MBB SED fits to Herschel
SPIRE data, after background subtraction. The red circle in frame a shows the
reference region used for the background estimation. Areas with low column
density have been masked in the plots (white pixels). The labelled stars indicate
the locations of embedded sources B1-B3. 
}
\label{fig:colden}
\end{figure*}

\subsection{Near-Infrared observations}

The central part of LDN~1642 has been imaged in the $J$, $H$, and $K_S$
bands using the HAWK-I instrument. HAWK-I is a cryogenic wide-field
NIR camera installed at the ESO VLT telescope
\citep{KisslerPatig2008}. The field of view is
7.5$^{\prime}\times$7.5$^{\prime}$, with a cross-shaped gap of
15$^{\prime\prime}$ between the four HAWAII 2RG 2048$\times$2048
pixel detectors. The pixel scale is 0.106$^{\prime\prime}$/pixel.
Further details of the instrument can be found in
\citet{KisslerPatig2008}. The observations were performed as ON-OFF
measurements to recover the faint surface brightness. The observations
consisted of three pointings arranged around the source B1, which was
too bright for direct observations. The NIR photometry was done using
the APPHOT task in Image Reduction and Analysis Facilities (IRAF)
software, and the final calibration was provided by the comparison
with the magnitudes in the 2MASS catalogue \citep{Skrutskie2006}. 

To study the surface brightness, we created another set of images 
where the stars were eliminated. To remove the stars, we first used
the DAOPHOT task ALLSTAR, which in many cases leaves significant
residuals at the location of bright stars. The effect of some bright
stars extends beyond the area masked by DAOPHOT. In such cases, the
masks were extended manually, removing areas where the surface
brightness enhancement was visibly above the general background.
Finally, faint stars that were not identified by DAOPHOT were
removed with median-filtering. The size of the median filter was
5.0$^{\prime\prime}$ (or 19 pixels) for $J$, $H$, $K_S$, and WISE 3.4\,$\mu$m
bands.  

For the $J$- and $K_S$-bands we calculated estimates of the absolute sky
brightness behind the LDN~1642 cloud by subtracting from DIRBE measurements the
combined flux of 2MASS stars, weighted by the DIRBE beam. This procedure
gave 71\,kJy\,sr$^{-1}$ and 31\,kJy\,sr$^{-1}$ for the $J$ and $K_S$ bands,
respectively. A linear interpolation gives 53\,kJy\,sr$^{-1}$ for the $H$ band.
The values have considerable uncertainty since the estimates vary  by
$\sim$30\% when derived from neighbouring independent DIRBE pixels. When
combining the DIRBE and 2MASS data, we do not explicitly include colour
corrections that are small compared to this uncertainty \citep{Levenson2007}.

\subsection{Optical observations}

The optical data were obtained from the Infrared Science Archive
(IRSA)\footnote{\url{https://irsa.ipac.caltech.edu/data/DSS/}} and are part of
the Space Telescope Science Institute (STScI) Digitized Sky Survey (DSS).
LDN~1642 was imaged with the UK Schmidt telescope to a photographic plate
(emulsion type IIIaF) using an OG590 filter, which roughly corresponds to the $R$
band, covering the wavelength range 6300-6900\,\AA. The plate covers a $\sim$
$6^{\circ} \times 6^{\circ}$ area on the sky and the digitised images have a
pixel size of 1$\arcsec$.  The DSS data are shown in
Fig.~\ref{fig:DSS_area}. The conversion to units of MJy\,sr$^{-1}$ is described
in Appendix~\ref{app:DSS}.

There are two bright stars, HD29613 (red giant) and HD29503 (1~Eri, a
triple star), at a small angular distance from LDN~1642. With the
parallax measurements in the \textit{Gaia} \citep{Gaia2016} Data
Release 2 catalogue \citep[DR2][]{GaiaDR2} (see also
\citet{Bailer-Jones2015, Luri2018}) we estimate $\sim 65$\,pc and
$\sim 36$\,pc as the distances of the two sources, respectively. The
cloud is at a more than 70\,pc larger distance, and the stars are thus
not likely to contribute significantly to its surface brightness.

\begin{figure*}
\includegraphics[width=17.5cm]{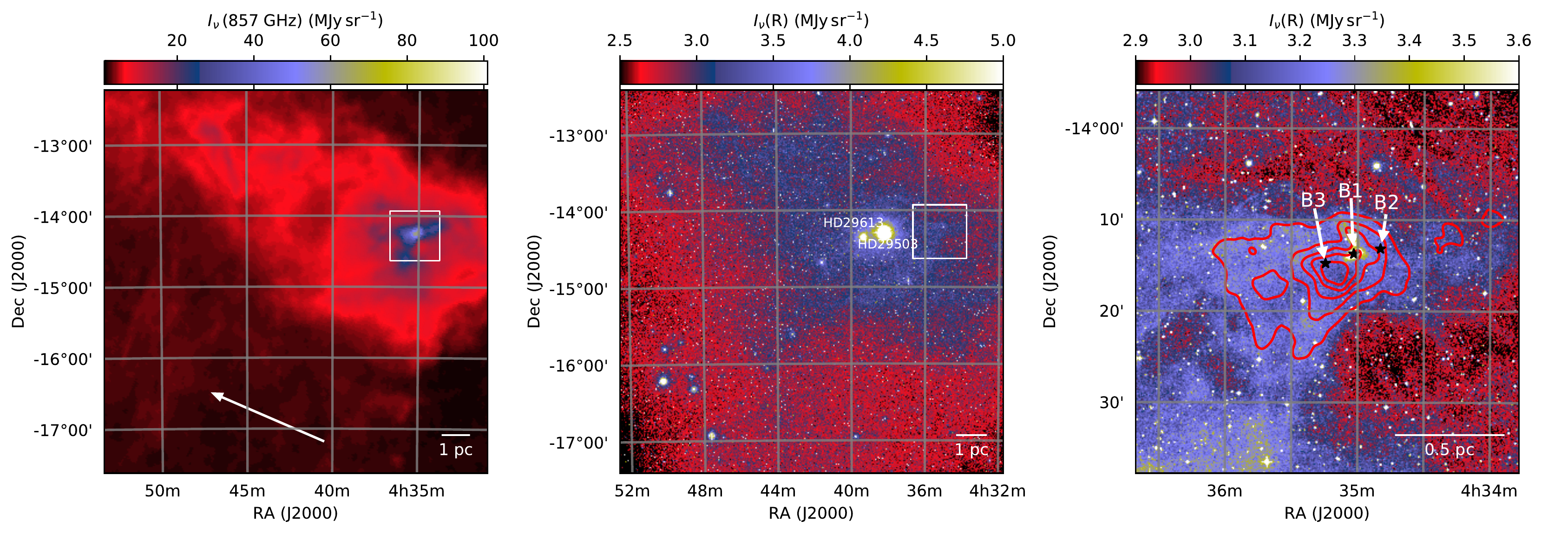}
\caption{
Large-scale environment of LDN 1642 shown by dust emission and optical
scattered light. Left panel shows the Planck 857\,GHz surface
brightness, the white arrow indicating the direction towards the
Galactic Plane. Centre panel shows the optical scattered light from
DSS red channel. The white box shows the location of the LDN~1642
cloud and covers an area of $0.7^\circ \times 0.7^\circ$. 
%% The image has been smoothed by a $3 \times 3$ Gaussian beam. 
Right panel: A close-up of the area marked with the white box in the
centre panel, showing the extended $R$-band surface brightness.
%% smoothed by a $3 \times 3$ Gaussian beam. The photon density has been truncated
%% at 7500. 
The black stars indicate the locations of the embedded sources B1, B2, and B3,
and the red contours show the $N(\rm H_2)$ column density. The lowest contour is
at $15 \, \%$ of the peak column density of $5.14 \times 10^{21}$ cm$^{-2}$, 
with the contour levels increasing in  $15 \, \%$ intervals.}
\label{fig:DSS_area}
\end{figure*}

\section{Results} \label{sect:results}

\subsection{Column densities and extinction} \label{sect:colden}

Column densities and dust temperatures were estimated using the
Herschel 160-500\,$\mu$m surface brightness measurements.  The maps
were resampled onto common pixels and modified blackbody fits were
performed pixel by pixel. The dust opacity spectral index was fixed to
$\beta=1.8$ \citep{Planck2011_MC,GCC-VI} and the conversion from
optical depth to hydrogen column density assumes a dust opacity of
$\kappa_{\nu}=0.1$\,g\,cm$^{-2} \,(\nu/1000\,{\rm GHz})^{\beta}$
\citep{Beckwith1990}. This value of $\kappa_{\nu}$ corresponds to dust
properties in very dense regions and has been used in earlier analysis
of LDN\,1642 \citep[e.g.,][]{GCC-III}.

One set of calculations was done with 160-500\,$\mu$m maps convolved to a common
41$\arcsec$ resolution, providing dust colour temperature and column density maps
at the same resolution (Fig.~\ref{fig:colden}a). Another column density map was
calculated at a higher resolution following the procedure described in
\citet{Palmeirim2013}, combining estimates computed with 160-250\,$\mu$m,
160-350\,$\mu$m, and 160-500\,$\mu$m data. This procedure provides a
$N({\rm H}_2)$ map at the resolution of the 250\,$\mu$m observations that was
further smoothed to 25$\arcsec$ resolution.

Extinction maps were calculated using the combination of 2MASS
\citep{Skrutskie2006} data and our HAWK-I photometry. For the latter,
we included all stars with measured magnitudes at least in the $H$ and
$K_S$ bands. We used a variation of the NICER method
\citep{Lombardi2001} that takes into account the estimated ratio
between the average extinction within a resolution element and
extinction towards individual stars \citep{Juvela_2016_allsky}. These
ratios were read from the Herschel column density map. The resolution
of that map is higher than the resolution of the final extinction maps
(41$\arcsec$ vs. 2$\arcmin$), which helps to reduce the noise caused
by column density variations on scales below 2$\arcmin$.
Figure~\ref{fig:colden}c shows the resulting extinction map of
$\tau(J)$. The extinction peaks at $A_J=2.6$\,mag ($A_{\rm
V}=9.3$\,mag for $R_V=5.1)$).

\subsection{Comparison of sub-millimetre and near-infrared data}
\label{sect:opacity_ratio}

\begin{figure}
\includegraphics[width=8.8cm]{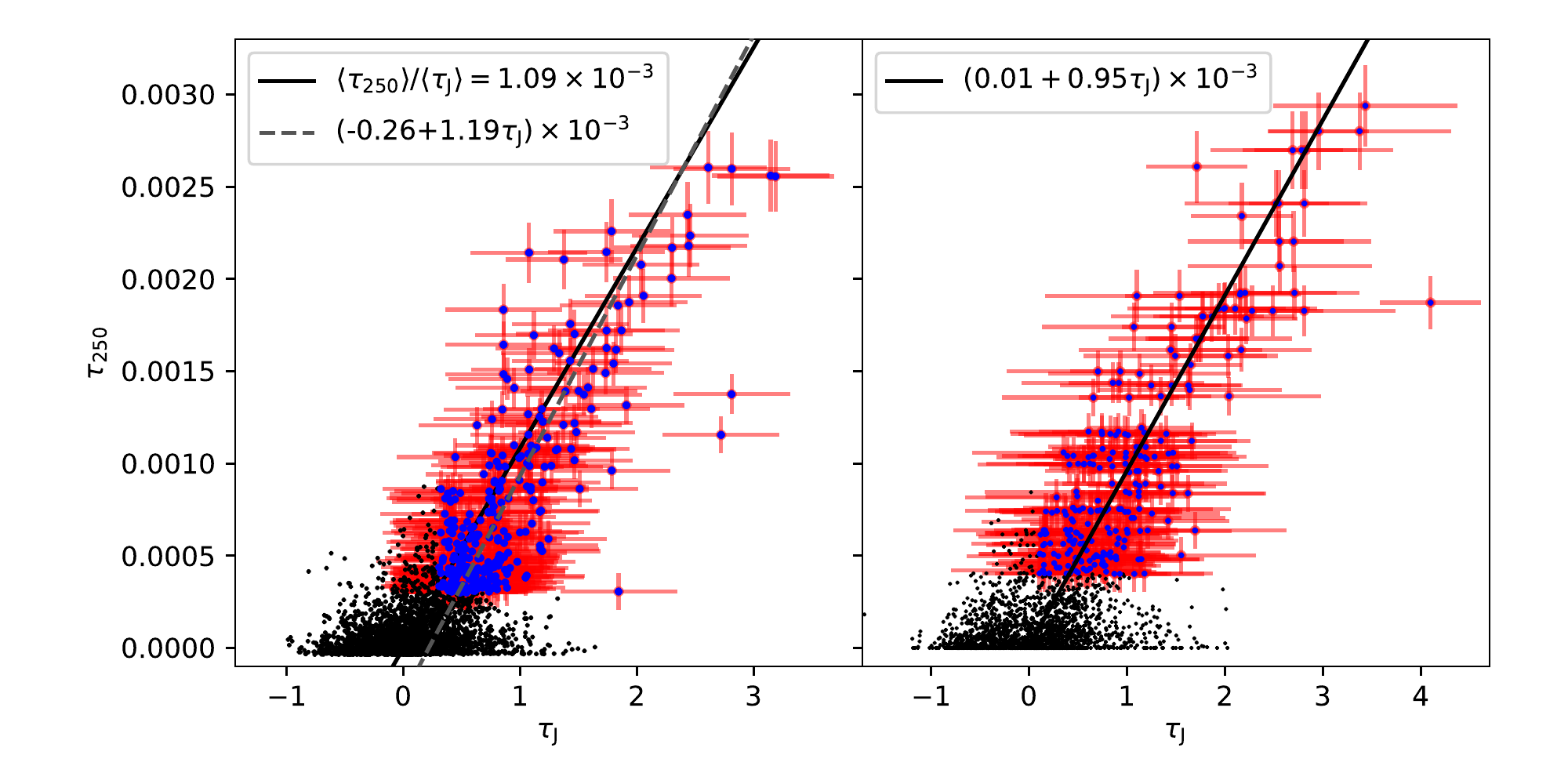}
\caption{
Correlations between the optical depths at 250\,$\mu$m and in the J
band. The left frame uses a NICER $\tau(J)$ map with 2$\arcmin$
resolution. 
The pixel values are plotted as black points. The blue points correspond to
the same data but excluding pixels close to the background level. 
The dashed line is the least-squares line to the blue
points and, for comparison, the solid line has a slope corresponding
to the ratio of the averages of the background-subtracted values. In
the right hand frame, we compare $\tau(J)$ estimates of
individual stars to the Herschel column density estimates at
40$\arcsec$ resolution. The parameters of the least-squares lines are
given in the frame.
}
\label{fig:tau_ratio}
\end{figure}

To quantify the ratio of sub-millimetre and NIR opacities, we
calculated the ratio $\tau(250\mu{\rm m})/\tau(J)$. The first
estimates are based on the NICER extinction map at 2$\arcmin$
resolution and the Herschel column density map convolved to the same
resolution. To establish a common zero point, the average value in
region defined by $\tau(250\mu{\rm m})<10^{-4}$ was subtracted from
both maps. With data remaining above $\tau(250\mu{\rm m})=10^{-4}$ and
sampled at 1 arcmin steps, we obtained a ratio $\langle \tau(250\mu
{\rm m})\rangle / \langle \tau(J)\rangle = 1.07\times 10^{-3}$.
Figure~\ref{fig:tau_ratio}a also shows a linear total least squares
fit to part of the data (blue points). The slope of the fit gives
$\tau(250\mu {\rm m})/ \tau(J) = (1.22 \pm 0.04)\times 10^{-3}$,
with the formal error estimates from the least squares fit. 

The above values are based on extinction maps at low resolution. For comparison,
we also correlated the extinction estimates of individual stars with the column
densities read from a map with 40$\arcsec$ resolution. This fit is shown in
Fig.~\ref{fig:tau_ratio}b. The result was $\tau(250\mu {\rm m})/ \tau(J) = (0.95
\pm 0.04)\times 10^{-3}$. Thus, with more conservative error estimates, the
optical depth ratio in LDN~1642 is $\tau(250\mu {\rm m})/ \tau({\rm J}) = (1.0
\pm 0.2)\times 10^{-3}$. Even this uncertainty may not fully cover all systematic
errors such as the possible bias in $\tau(250\mu {\rm m})$ caused by LOS
temperature variations.

\begin{figure}
\includegraphics[width=8.8cm]{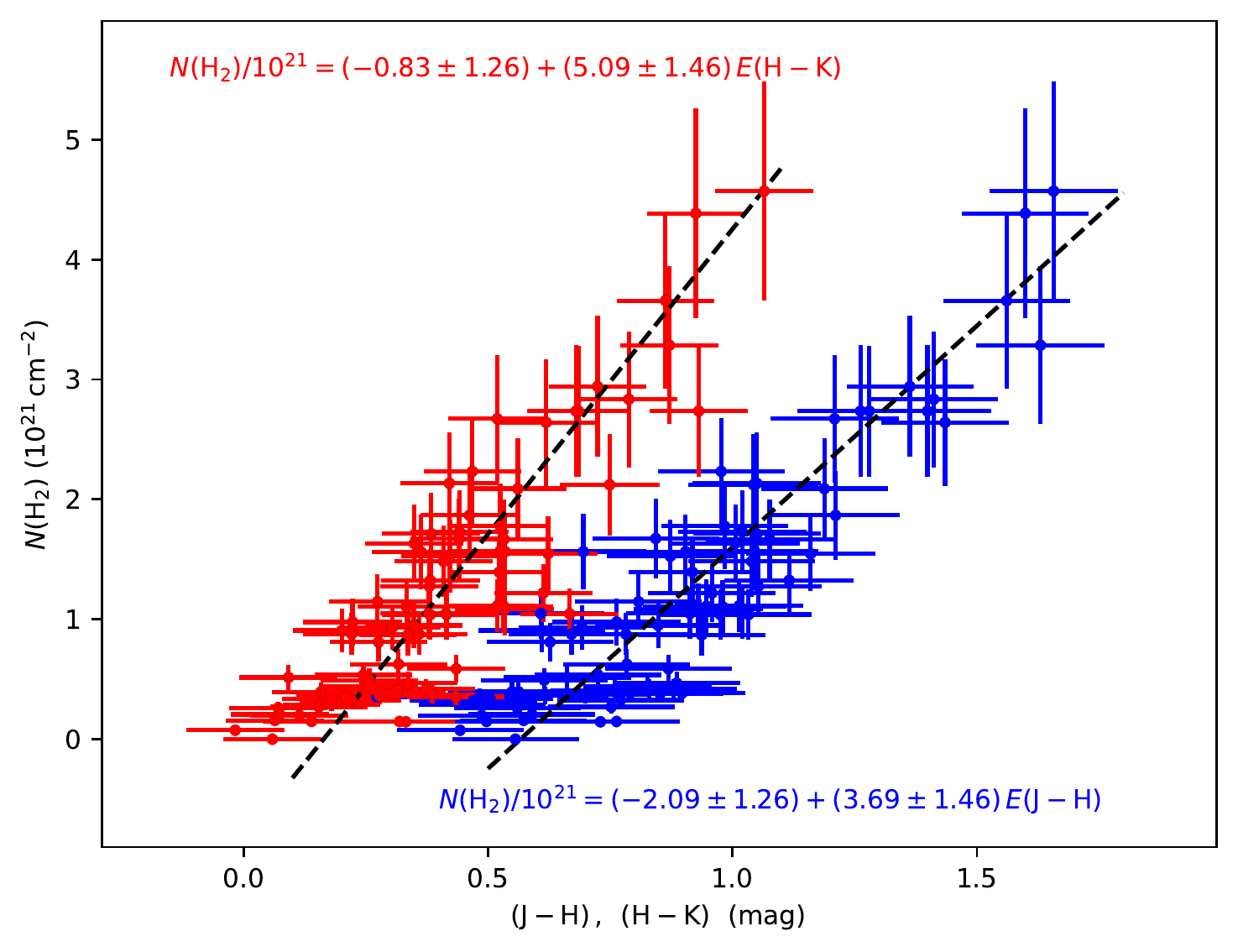}
\caption{
Estimated column density $N({\rm H}_2)$ as function of NIR colours
$J$-$H$ (in blue) and $H$-$K_S$ (in red). The parameters of the linear
fits are given in the plot.
}
\label{fig:extinction_curve}
\end{figure}

The derivation of $\tau(J)$, as shown in Fig.~\ref{fig:tau_ratio}, assumed
the standard extinction curve of \citet{Cardelli1989}. We also correlated the
$N({\rm H}_2)$ data at 25$\arcsec$ resolution directly with the NIR J-H and
H-K colours of individual stars (Fig.~\ref{fig:extinction_curve}). The
relationships remain linear up to the highest values and for both colour
excesses. 
%% are fitted using orthogonal line
%% regression\footnote{https://docs.scipy.org/doc/scipy/reference/odr.html}.
The $N({\rm H}_2)$ uncertainties are assumed to be 20\% and the
uncertainties of the NIR colours are the squared sums of the
photometric errors and a constant that represents the dispersion in
the intrinsic colours and is set so that the final estimates of
uncertainty are consistent with the observed scatter. The numerical
values quoted in the figure depend on the chosen value of
$\kappa(250\,\mu{\rm m})$, but the ratio of the slopes gives
independently $E({\rm H-K})/E({\rm J-H})=0.73\pm0.35$. This value is
slightly higher than in the \citet{Cardelli1989} extinction curve
(0.66), but the difference is not significant considering the
uncertainties. The result also would change by less than 1\% if $N({\rm
H}_2)$ was taken from the other column density map, which was derived
from the 250-500\,$\mu$m Herschel data at 41$\arcsec$ resolution. The
extrapolation of the linear relationships to zero column density shows
the average NIR colours in the region used for the $N({\rm H}_2)$
background subtraction, $\langle J-H \rangle$=0.57\,mag and $\langle
H-K \rangle = 0.17$\,mag.

% Observational results on scattered light
% comparison with dust emission

\subsection{Scattered light} \label{sect:scattered_light}

\begin{figure}
\centering
\includegraphics[width=8.8cm]{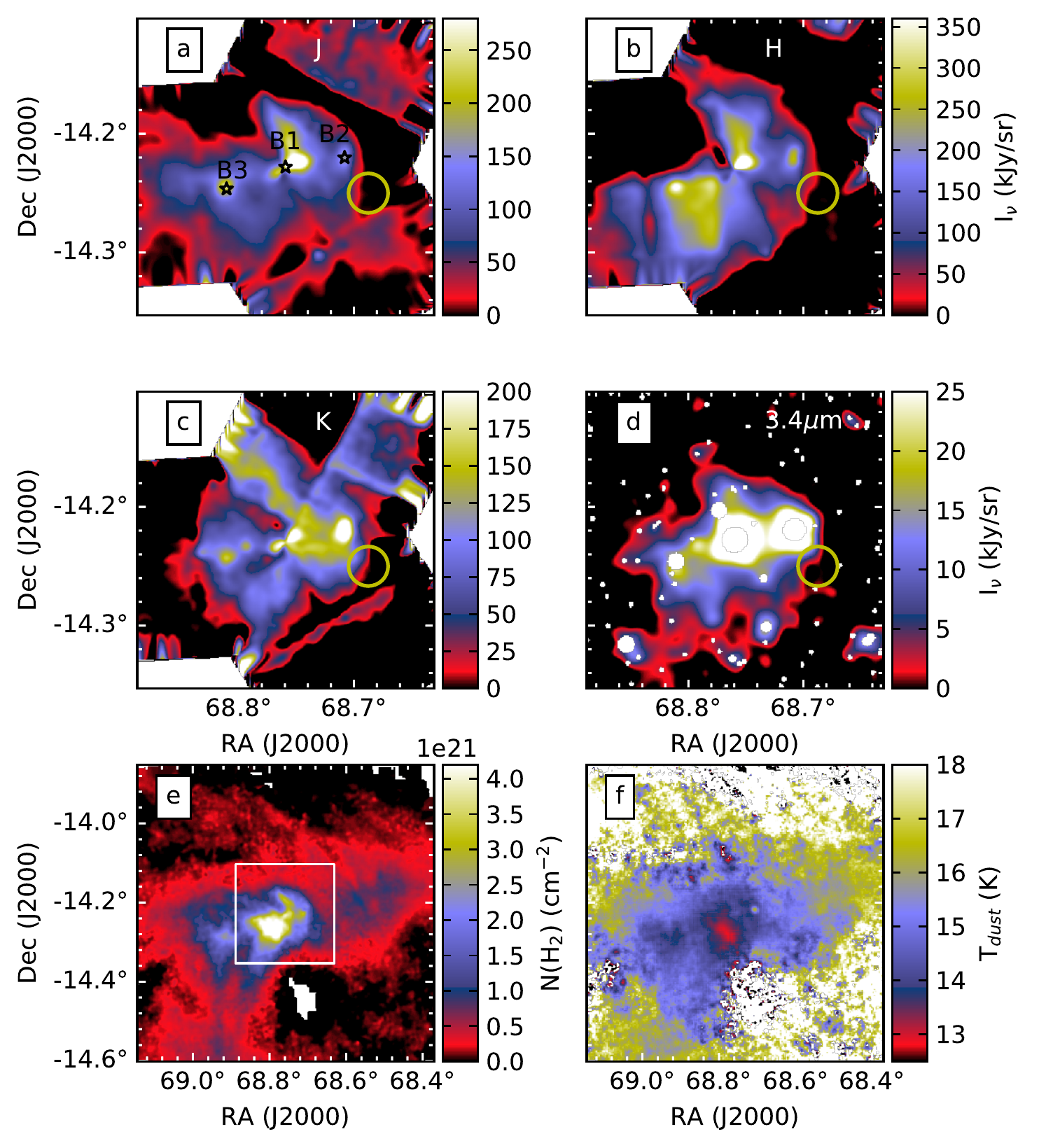}
\caption{%
Observed surface brightness in $J$, $H$, $K_S$, and WISE 3.4\,$\mu$m bands
(frames a-d, respectively). Herschel column density and dust
temperature maps are shown in frames e and f. Bright stars identified
by DAOPHOT were removed and replaced with interpolated surface
brightness, and faint stars have been eliminated with median
filtering. In the WISE 3.4\,$\mu$m band image (frame d), the white
regions correspond to areas around bright stars that were excluded
from subsequent analysis. The YSOs B1-B3, are identified in frame a.
The yellow circle in frames a-d indicate the reference region used for
setting a common zero level. The white box in frame e corresponds to
the extent of the $J$, $H$, $K_S$, and 3.4\,$\mu$m maps (the area shown in
frames a-d).
}
\label{fig:surfb}
\end{figure}

In this section we compare the column density estimated from sub-millimetre dust
emission with the surface brightness maps for NIR dust scattering. The final
surface-brightness maps of the central part of the LDN~1642 cloud in the $J$,
$H$, $K_S$, and WISE 3.4\,$\mu$m bands are shown in Fig. \ref{fig:surfb}, which
also includes for comparison the column density and dust temperature maps derived
from Herschel observations. All NIR maps have been convolved to the same
25$^{\prime\prime}$ resolution, to enable comparison between the NIR surface
brightness and column density. To establish a common zero point, we have
subtracted the median values of a region marked with the yellow circles in Fig.
\ref{fig:surfb}. The surface brightness values for $J$, $H$, and $K_S$ are
similar within a factor of two, while for WISE 3.4\,$\mu$m, the values are almost
ten times lower. LDN~1642 contains a central dense region surrounded by diffuse
material. We have studied the densest part of the cloud, which contains three
YSOs, B-1, B-2, and B-3. These three YSOs are detectable in $J$, $H$,
$K_S$, and 3.4\,$\mu$m maps as compact sources. North of B1, an elongated
structure is visible in surface brightness and Herschel column density maps. It
is more prominent in $J$ and $H$ compared to the longer wavelengths, $K_S$ and
3.4\,$\mu$m.

\begin{table*}
\caption{Comparison between Herschel column density map with observed
surface brightness. Columns are: (1) correlated quantities, (2) slope of linear
fit, (3) intercept of linear fit, (4) Spearman's correlation
coefficient $r$, and (5) median value of $I_{\nu}$/$N$(H$_2$) ratios.
\label{tbl:JHKW_NH2_comp}}
\centering
%%\resizebox{1.0\linewidth}{!}{
\begin{tabular}{lcccc}
  \hline\hline
  Quantity  & Slope & Intercept & $r_s$ & Ratios \\
  & ($\rm kJy\,sr^{-1}\,cm^{2}$) & ($\rm kJy\,sr^{-1}$) & & 
  ($\rm kJy\,sr^{-1}\,cm^{2}$) \\
  (1)           & (2) & (3) & (4) & (5)  \\
  \hline
  \multicolumn{2}{l}{Smaller region}  \\
  $I_J$ $-$ $N$(H$_2$) & $1.00\times 10^{-19}$ & -17.2 & 0.97 & $8.28\times 10^{-20}$ \\
  $I_H$ $-$ $N$(H$_2$) & $8.46\times 10^{-20}$ &  61.3 & 0.86  & $1.35\times 10^{-19}$ \\
  $I_K$ $-$ $N$(H$_2$) & $2.24\times 10^{-20}$ &  82.9 & 0.47  & $8.62\times 10^{-20}$ \\
  $I_{3.4\,\mu{\rm m}}$ $-$ $N$(H$_2$) & $9.20\times 10^{-21}$ & 0.614 & 0.91 & $7.59\times 10^{-21}$ \\
  \hline
  Larger region  \\
  $I_J$ $-$ $N$(H$_2$) & $1.99\times 10^{-20}$ & 39.5 & 0.74 & $5.72\times 10^{-20}$ \\
  $I_H$ $-$ $N$(H$_2$) & $8.61\times 10^{-20}$ & 35.6 & 0.83 & $7.09\times 10^{-20}$ \\
  $I_K$ $-$ $N$(H$_2$) & $1.79\times 10^{-20}$ & 31.1 & 0.40 & $3.62\times 10^{-20}$ \\
  $I_{3.4\,\mu{\rm m}}$ $-$ $N$(H$_2$) & $3.72\times 10^{-21}$ & 2.875 & 0.60 & $4.34\times 10^{-21}$ \\
\hline
\end{tabular} 
%% }
\end{table*}

%% \subsubsection{Comparison between surface brightness and column density estimates}

\begin{figure}[h]
\centering
\resizebox{9.0cm}{8.5cm}{\includegraphics{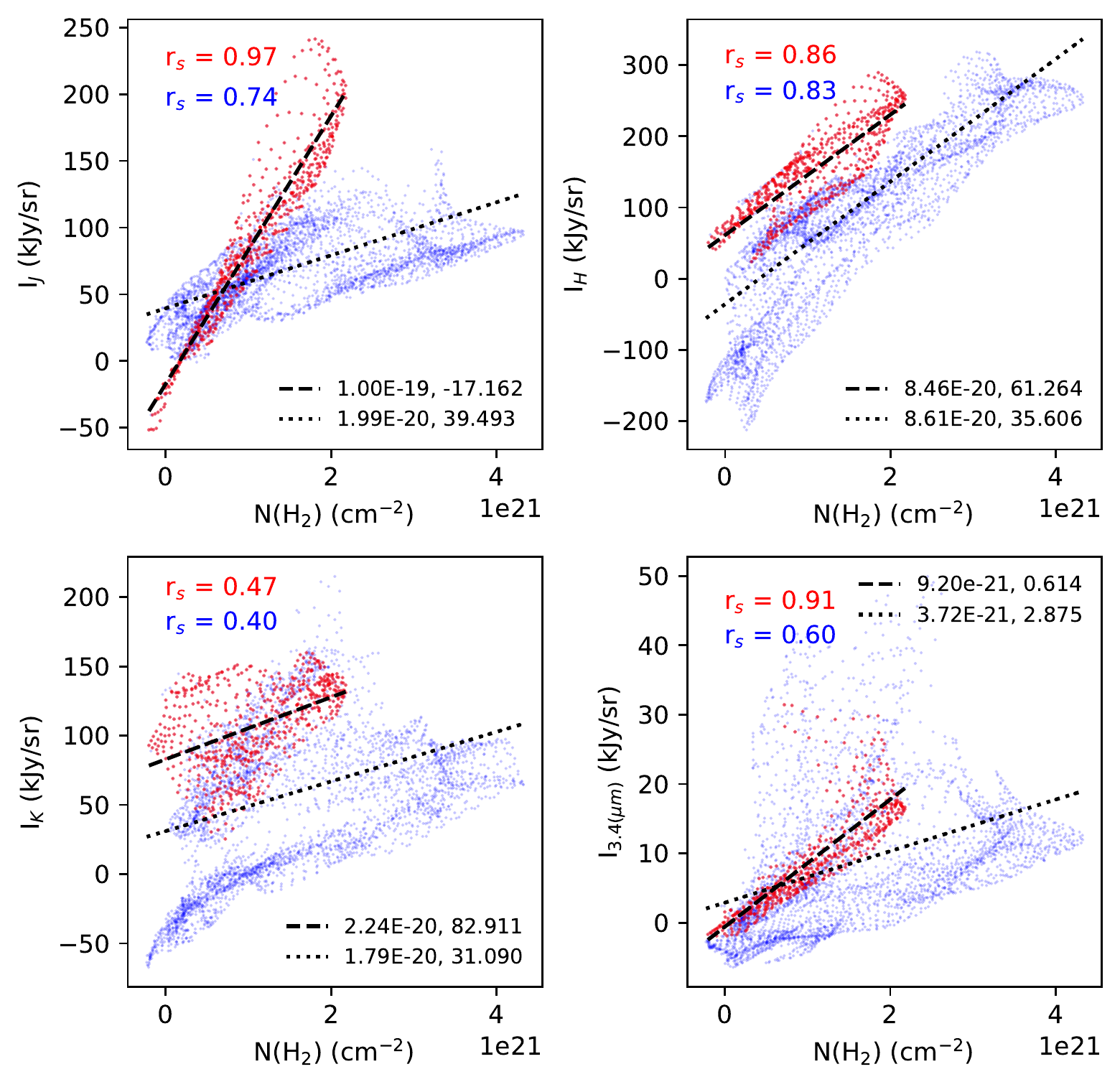}}
\caption{
Correlation between the observed scattered light in the $J$, $H$, $K_S$, and
3.4\,$\mu$m bands and Herschel column density. The data are at
25$^{\prime\prime}$ resolution and sampled at 7$^{\prime\prime}$ steps
from the two areas marked in Fig. \ref{fig:cmp_JHK_model_all}f, blue
points for the larger and red points for the smaller area. The dashed
lines and the dotted lines are robust linear least squares lines
fitted to the red points and blue points, respectively. The fitted
slopes and intercepts are quoted in the frames. The Spearman's
correlation coefficients are also given in the corresponding colours.
}\label{correlation_JHKW}
\end{figure}

Sub-millimetre dust emission shows good correlation with the morphology of NIR
emission as shown in Fig. \ref{fig:surfb}. The correlation between the observed
scattered light in the $J$, $H$, $K_S$, and 3.4\,$\mu$m bands with Herschel
column density is shown in Fig.~\ref{correlation_JHKW}. To make a pixel-to-pixel
comparison of observed $J$, $H$, $K_S$ and 3.4\,$\mu$m surface brightness with
Herschel column density maps, we selected two regions (marked in
Figs.~\ref{fig:plot_scattered_light}e and \ref{fig:cmp_JHK_model_all}f). Within
the larger region, we masked the positions of the three YSOs, B-1, B-2 and B-3 to
reduce their effect.  The maps at 25$^{\prime\prime}$ resolution are sampled at
7$^{\prime\prime}$ steps. In Fig.~\ref{correlation_JHKW}, the blue points show
the comparison for the larger area and the red points for the smaller area. We
fitted robust linear least squares lines for both regions. The observed NIR
surface brightness shows strong correlation with Herschel column density in all
the bands.
%% We also estimated the Spearman's correlation coefficients for all the bands. 
The slope decreases for both regions as the wavelength increases from
$J$-band to 3.4\,$\mu$m (Fig. \ref{correlation_JHKW}). Table
\ref{tbl:JHKW_NH2_comp} lists the parameters estimated from the
comparison between the Herschel column density map and the surface
brightness maps. The correlation coefficients indicate strong
correlation in the $J$, $H$ and 3.4\,$\mu$m bands and moderate
correlation in the $K_S$ band.

\section{Radiative transfer modelling}   \label{sect:RT}

We constructed radiative transfer (RT) models for the dust emission to derive 3D
models of the density distribution in LDN1642. Next, the obtained density
field and assumptions of the external and internal radiation sources were used to
calculate predictions for the scattered light.

\subsection{Modelling of Herschel emission} \label{sect:mod_Herschel}

A 3D density model of LDN1642 was first optimised to match the Herschel
250-500\,$\mu$m observations. The cloud volume was divided to 144$^3$ cells, each
with a linear size of $\sim0.0136$\,pc. This length scale corresponds to
20$\arcsec$ at the distance of 140\,pc. The model was further refined according
to the local density by adding up to two levels in the octree hierarchy, with an
approximately equal number of cells on each of the three refinement levels. The
model thus reaches a resolution of $5.0\arcsec$ over most of the dense areas.

The LOS density profile corresponding to a map pixel (along the third dimension) was set equal to
the narrowest column density profile that existed for any line (any position angle) crossing that pixel
on the plane of the sky. This default LOS profile is in the following referred to as having a relative
width of $W=1$. It favours a cylindrical 3D geometry for features that appear elongated on the observed
surface brightness maps. Because of the fundamental difference between the density and projected
column density profiles, this setup corresponds only approximately to cylinder symmetry.
Furthermore, because the actual LOS extent of the cloud and the inclination of the structures are
unknown, we tested for comparison models with 70\% larger LOS extent ($W=1.7$).

The angular distribution of the incoming radiation was obtained from the DIRBE
all-sky maps \citep{Hauser1998}, which include direct observations of the $J$ and
$K_S$ bands. At shorter wavelengths, the angular distribution is assumed to be
the same as in the $J$ band. This assumption is not entirely accurate but
includes the main effects of the anisotropic external illumination. The spectrum
of the external radiation field was rescaled to match the \citet{Mathis1983}
estimates of the radiation field intensity in the solar neighbourhood. In the NIR
regime, the \citet{Mathis1983} estimates are some 40\% below the sky-averaged
DIRBE values.

For the first models, the dust properties were taken from \cite{Jones2013} (in
the following J13). In dense clouds, such as LDN~1642, the dust is expected to
evolve towards larger grain sizes and larger sub-millimetre opacity. To quantify
the potential effects of this dust evolution, as a second option we tested models
that contained only ice-coated aggregate grains (AMMI). A third set of model
clouds was also created using spatial variations in the relative abundance of J13
dust, core-mantle-mantle (CMM) grains, and AMMI grains. In the following, these
models are referred to as THEMIS models. \citet{Ysard2016} have already
used CMM and AMMI dusts to model enhanced NIR and MIR scattering (cloudshine and
coreshine).  In the THEMIS cloud models, the relative abundances of the three
dust components were set according to the function
\begin{equation}
x = 0.5[(1+\tanh(2\log (\frac{n}{n_1})))-(1+\tanh(2\log
(\frac{n}{n_2})))].
\end{equation}
The threshold density values ($n_1$, $n_2$) were set equal to (1, $10^4$),
$(10^4, 4\times 10^4)$, and $(4\times 10^4, 10^{10})$\,cm$^{-3}$ for J13, CMM,
and AMMI dust, respectively.  Thus, the low-density parts of the model
cloud have only J13 dust. Its abundance drops to zero around $10^4$\,cm$^{-3}$,
where CMM is briefly the most abundant component before the relative abundance of
AMMI rises from zero to one at densities above $4\times 10^4$\,cm$^{-3}$. 

The calculations were performed with the Monte Carlo RT programme SOC
\citep{Juvela2019_SOC}, assuming that sub-millimetre emission can be
predicted with calculations where the grains remain at an equilibrium
temperature. Because of the associated larger computational cost, full
calculations with stochastically heated grains were performed, for
reference, only in one case (J13 dust, $W=1$).

The model predictions for surface brightness were saved as $288\times 288$ pixel
maps with a pixel size of 5$\arcsec$. To match the model intensities with
the Herschel observations, the models were optimised iteratively.
The column densities were updated using the ratio of the observed and
the model-predicted 350\,$\mu$m surface brightness values. This gave
for each map pixel a scaling factor that was used to update the
densities in all cells along the corresponding LOS.
The external radiation field was adjusted with a single scalar factor equal to
the average 250\,$\mu$m to 500\,$\mu$m intensity ratio in observations, divided
by the same ratio in the model predictions. This approach converges the
average colour temperature of the model towards the observed average
250-500\,$\mu$m colour temperature. In addition to the external radiation field,
we included point sources at the locations of the three known embedded sources
(see Fig.~\ref{fig:colden}). These were modelled as T=6000\,K black bodies and
their luminosities adjusted so that the model predictions for the average dust
colour temperature in a small $5\times5$ pixel area around the sources matched
the observations. The heating from the embedded sources reduces the predicted
column densities but only in a very limited area. When the scattered light is
later modelled in the neighbourhood of the embedded sources, the density field
predicted by this emission modelling is not used (Sect.~\ref{sect:PS_models}).

\begin{figure*}
\sidecaption
\includegraphics[width=12cm]{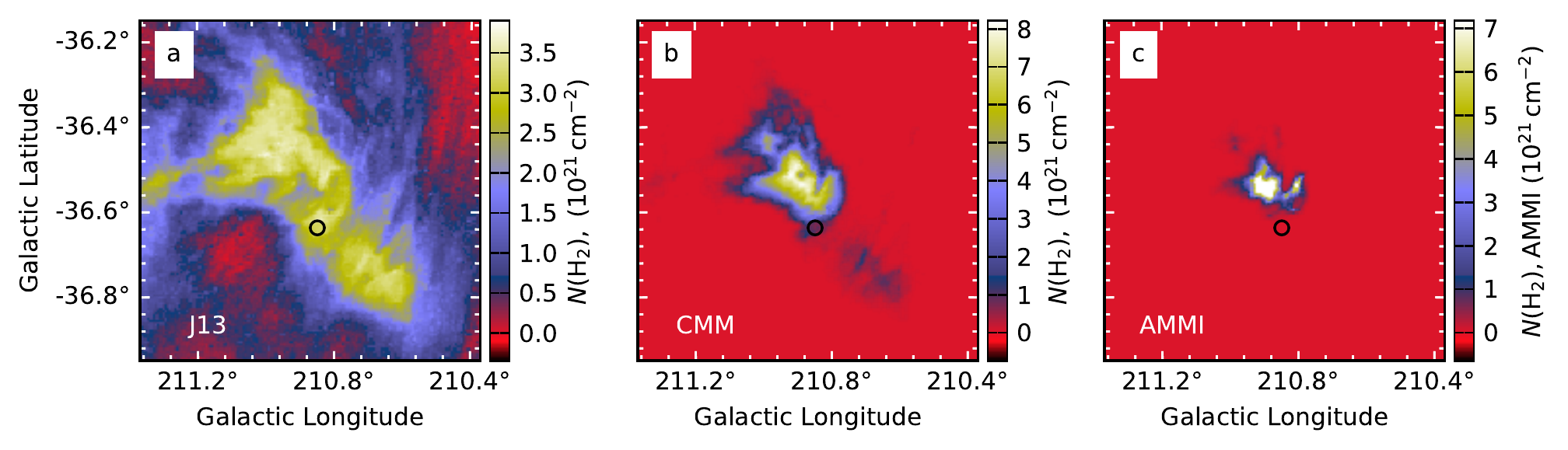}
\caption{
Hydrogen column density $N({\rm H}_2)$ associated with each dust
component in the THEMIS model with spatially varying dust abundances.
The three frames correspond to the J13, CMM, and AMMI dust,
respectively. The black circle shows the reference area used in
scattered-light analysis (cf Fig.~\ref{fig:surfb}).
}
\label{fig:abundance_colden}
\end{figure*}

The comparison of the observed and model-predicted surface brightness
maps is shown in Appendix~\ref{app:emission}. The emission is fitted
almost equally well using any of the dust models and with both cloud
shapes, $W=1$ and $W=1.7$.  Figure~\ref{fig:abundance_colden}
shows the abundance variations in the THEMIS model, by plotting the
column densities weighted by the relative amounts of the J13, CMM, and
AMMI dust.

The emission models predict NIR optical depths $\tau(J)$ that are above NICER
estimates, i.e. compared to direct extinction measurements based on background
stars. To quantify this difference, we convolved the $\tau(J)$ model maps
to the 2$\arcmin$ resolution of the NICER map and subtracted from both the
average value in the area where $\tau(J)$ of the J13 model was within the 1-5\%
percentile range. The ratios of the average model-predicted and observed
$\tau(J)$ were then calculated for the pixels falling in the 25-90\% percentile
range of the J13 map. The lower limit was chosen so that the comparison avoids
regions where the signal is close to zero, and the upper limit of 90\% was
chosen to downweight the contribution of the cloud centre, where the low
stellar density renders the NICER estimates more uncertain. However, the obtained
optical-depth maps were quite flat, without systematic variations correlated with
the column density.

\begin{table}
\caption{%
NIR optical depths of model clouds ($\tau_J^{\rm M}$) relative
to NICER measurements ($\tau_J^{\rm N}$). The last column gives
$k_{\rm ISRF}$, the relative radiation field strengths of the models.
}
\begin{tabular}{lcccc}
\hline \hline
Model    & $\langle \tau_J^{\rm M}  \rangle$  &
$\langle \tau_J^{\rm N} \rangle$  &  
$\langle \tau_J^{\rm M}  \rangle$ / $\langle \tau_J^{\rm N} \rangle$  &
$k_{\rm ISRF}$ \\
\hline
J13, $W$=1.0    &  0.40    &  0.17   &   2.40  &  0.80  \\
J13, $W$=1.7    &  0.39    &  0.17   &   2.32  &  0.74  \\
%            J13X_W1.00   &  0.19  &  0.17  &   1.15  &  1.2118 \\
%         J13_W1.00_SHG   &  0.51  &  0.17  &   3.05  &  1.1393 \\
%    J13_W1.00_TAUJ0.26   &  0.38  &  0.17  &   2.28  &  1.8781 \\
%COM,  $W$=1.0   &     0.43  &  0.16  &   2.60  &  0.73 \\
%COM,  $W$=1.7   &     0.41  &  0.16  &   2.52  &  0.69 \\
%CMM,  $W$=1.0   &     0.72  &  0.15  &   4.67  &  0.59 \\
%CMM,  $W$=1.7   &     0.69  &  0.15  &   4.46  &  0.55 \\
AMMI, $W$=1.0   &     0.67  &  0.15  &   4.58  &  0.89 \\
AMMI, $W$=1.7   &     0.65  &  0.15  &   4.46  &  0.85 \\
%%              &     0.51  &  0.16  &   3.10  &  1.0648 \\
%%              &     0.41  &  0.16  &   2.48  &  1.6368 \\
%COM,  $W$=1.0, Shg\tablefootmark{a} &  0.51  &  0.16  &   3.10  &  1.06   \\
%COM,  $W$=1.0, Ext\tablefootmark{b} &  0.41  &  0.16  &   2.48  &  1.64 \\
THEMIS, $W$=1.0  &   0.45  &  0.17  &   2.69  &  0.78 \\  % THEMIS2 !!
%          THEMIS_W1.00   &  0.45  &  0.17  &   2.69  &  0.7822 \\
%         THEMISX_W1.00   &  0.20  &  0.17  &   1.17  &  1.2076 \\
%         THEMIS2_W1.00   &  0.45  &  0.17  &   2.69  &  0.7821 \\  ****
%        THEMIS2X_W1.00   &  0.20  &  0.17  &   1.17  &  1.2077 \\
J13,  $W$=1.0, Shg\tablefootmark{a}  &  0.51  &  0.17  &   3.05  &  1.14 \\
J13,  $W$=1.0, Ext\tablefootmark{b}  &  0.38  &  0.17  &   2.28  &  1.88 \\
\hline 
\end{tabular}
\tablefoot{ 
\tablefoottext{a}{Full calculations with stochastically
heated grains} 
\tablefoottext{b}{Assuming radiation field that is
attenuated corresponding to $\tau_J^{\rm Ext}=0.26$ } 
}
\label{table:tau_J}
\end{table}

The results are listed in Table~\ref{table:tau_J}. The NICER estimates are always
calculated using the extinction curve of the corresponding dust model, although
these change the results only a little. The NIR optical depths of the optimised
model clouds are 2.3 to 4.6 times higher than observed. The differences
depend mainly on the dust model. The LOS cloud extent has effects only at a level
of a few per cent. The full treatment of stochastic grain heating increased the
$\tau(J)$ value by only 25\% and the estimate of the radiation field
intensity by some 40\%. Qualitatively, the effect of stochastic heating is
expected to be similar for the other dust models. The three-component THEMIS
model is in this comparison close to the J13 model, because the J13 dust
component is dominant at large scales. The relative abundance of the dust
components is illustrated in Fig.~\ref{fig:abundance_colden}.

The differences in the predicted volume density and dust temperature 
distributions will be discussed in Sect.~\ref{dis:RT_emission}. We
will return to the optical depth discrepancy between the emission
models and the NICER maps in Sect.~\ref{sect:dis_scattering}.

\subsection{Modelling of extended scattered light} \label{sect:extended}

\begin{figure}
\includegraphics[width=8.8cm]{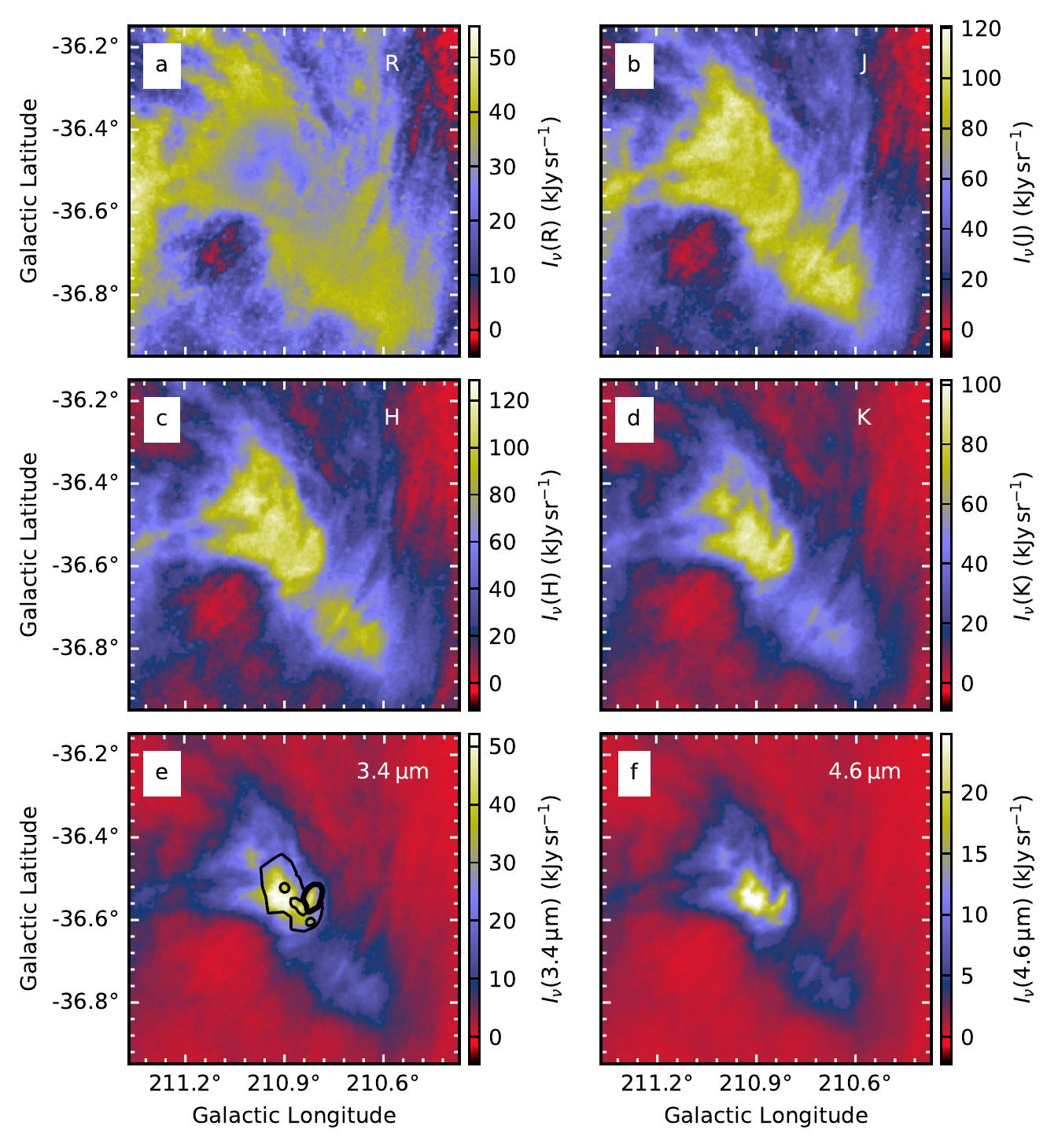}
%
%\sidecaption
\caption{
Predicted intensity of scattered light. The cloud model is J13 with
$W=1$. Frames a-f correspond to the bands R, $J$, $H$, $K_S$, 3.4\,$\mu$m, and
4.6\,$\mu$m, respectively. Only the scattering of the external
radiation field is included and the effects of attenuated sky
background are not included. 
Frame e indicates areas used for correlation analysis, the larger area
drawn with thin line (excluding the sources B1-B3) and the smaller
area with thick line (see also Fig.~\ref{fig:cmp_JHK_model_all}).
}
\label{fig:plot_scattered_light}
\end{figure}

We calculated predictions for the scattered light in the $R$, $J$, $H$, $K_S$,
and 3.4\,$\mu$m bands, using the density fields obtained from the emission
modelling (Sect.~\ref{sect:mod_Herschel}). Figure~\ref{fig:plot_scattered_light}
shows the results for one of the cases (J13, $W=1$). The figure includes the
4.6\,$\mu$m band although, for the lack of signal in the actual observations,
this wavelength will not be analysed any further. The figure only shows the
scattered light and thus does not include the effects of the LOS sky background,
which reduces the observed ON-OFF signal by $I_{\rm bg}(e^{-\tau}-1)$.

The $R$-band results are compared to DSS data in
Appendix~\ref{app:DSS}. We do not have an estimate for the absolute
brightness of the sky background in the $R$ band and therefore only
show the comparison with the scattered-light component from the
model. That emission is found to be only a fraction of the observed signal and
a positive sky background $I_{\rm bg}$ would decrease the model
predictions further. The $R$-band data are not analysed further in
this paper.

\begin{figure*}
\includegraphics[width=18.0cm]{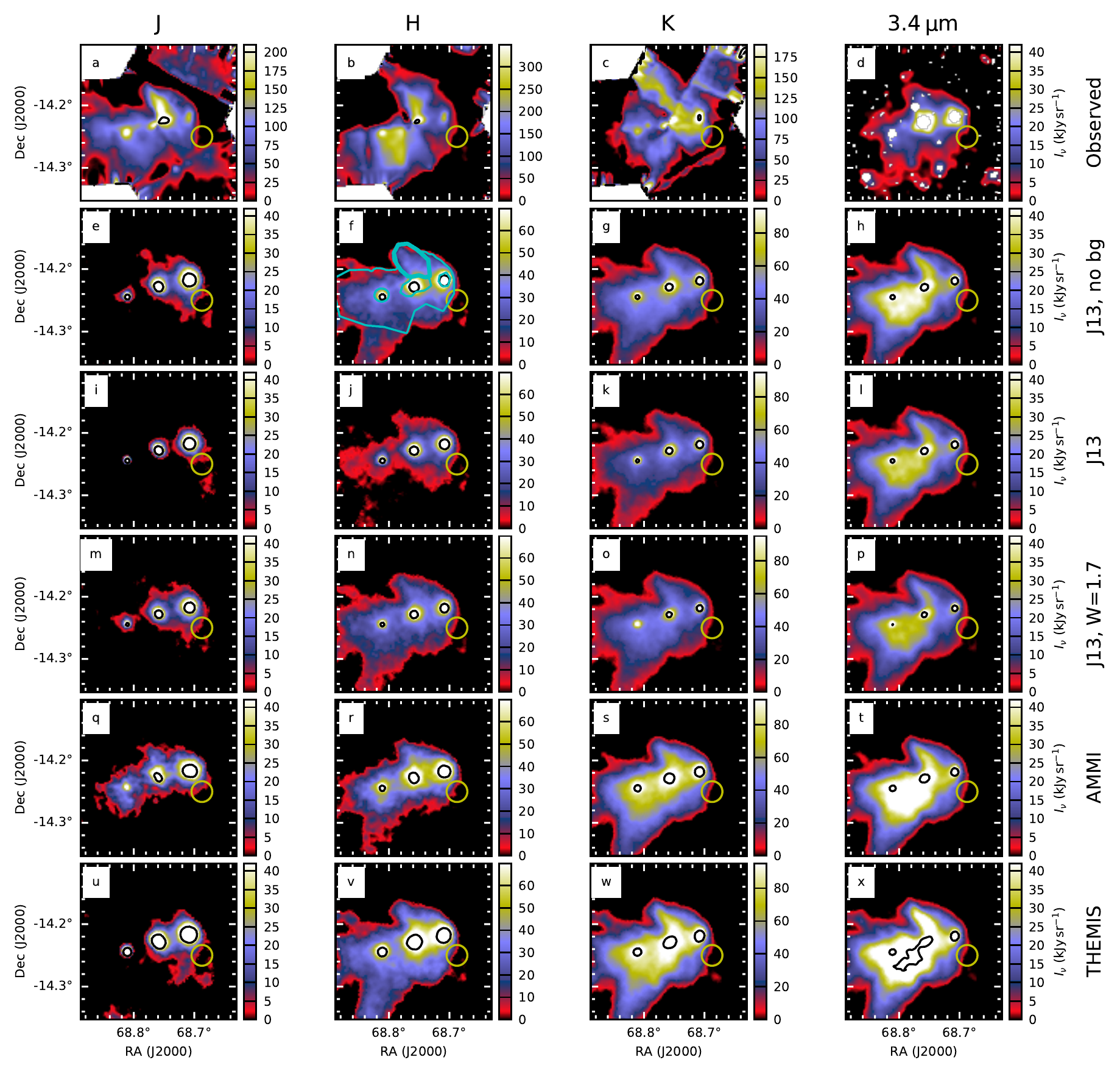}
%
%\sidecaption
\caption{
Comparison of observed (first row) and modelled (other rows)
1.25-3.4\,$\mu$m surface brightness. The second row corresponds to J13
dust, $W=1$, and no sky background ($I_{\rm bg}=0$). The effects of 
sky background are included in the other cases (J13 dust with $W=1$
and $W=1.7$, and AMMI with $W=1$, and the THEMIS model with three dust
components).
Frame f indicates the areas used for correlation analysis, the larger
area shown with a thin cyan line (excluding the sources B1-B3) and the
smaller area with a thick cyan line. 
The yellow circles indicate the reference region used for setting a
common zero level. Black contours are drawn at 1.5 times the maximum
of the colour scale. The colour scale is the same for all models but
different for observations.
}
\label{fig:cmp_JHK_model_all}
\end{figure*}

Figure~\ref{fig:cmp_JHK_model_all} compares the model predictions at $J$,
$H$, $K_S$, and 3.4\,$\mu$m with the corresponding HAWK-I and WISE observations.
After median filtering, the observed maps have been convolved to 25$\arcsec$
resolution. The model predictions have a similar resolution because the density
field was fitted to Herschel 350\,$\mu$m data that have $\sim 26\arcsec$
resolution (Sect.~\ref{sect:mod_Herschel}). This resolution applies to the
scattering of the external ISRF. The direct radiation from the embedded sources
and their scattered radiation are convolved with FWHM=25$\arcsec$ Gaussian,
although the effective resolution of the latter is not well defined. Near the
point sources, the surface brightness variations are dominated by the radial
change of the radiation field, which is to some extent resolved at a resolution
higher than that of the underlying density field. However, we will examine the
environment near the embedded sources separately in Sect.~\ref{sect:PS_models}
and here concentrate on the larger scales. At distances larger than $\sim
1\arcmin$ from the sources B1-B3, the scattering is dominated by the external
ISRF. To establish a common zero point for the observations and the models, we
subtract from each map the median value of the reference area indicated in
Fig.~\ref{fig:cmp_JHK_model_all}.

Figure~\ref{fig:cmp_JHK_model_all} includes model predictions for five cases. The first three are for
J13 dust, the first one showing the scattered light without the effect of the background sky brightness
$I_{\rm bg}$ on the ON-OFF measurement. The background term $I_{\rm bg}(e^{-\tau}-1)$ is included
in all other cases, clouds with $W=1.0$ and $W=1.7$ and with J13 dust, and the $W=1.0$ cloud with AMMI
and THEMIS dust cases. The predicted surface brightness is significantly below the observed levels and
is even negative for the $J$ band. The negative values result from the significant sky
brightness, combined with the significant NIR optical depth of the model clouds.

Because of the tension between the $\tau(J)$ values of the NICER
measurements and the cloud models that fit the dust emission, we
repeated the scattering calculations using modified model clouds
where the average $\tau(J)$ was decreased to the level of the
observed NICER values. This was done by dividing all densities
by the factors $\langle \tau_J^{\rm M} \rangle$ / $\langle
\tau_J^{\rm N} \rangle$ of Table~\ref{table:tau_J}. The results are
shown in Fig.~\ref{fig:cmp_JHK_model_all_rescaled}. The intensity of
the J13 model is still far too low but the pure AMMI model rises
almost to within a factor of two of the observed surface brightness
values.
%% ???
%%Fig.~\ref{fig:cmp_JHK_model_corr_rescaled} shows correlations for CMM
%%dust ($W=1$), where also the effects of the background sky brightness
%%$I_{\rm bg}$ are taken into account. 
The interpretation of these results will be discussed in
Sect.~\ref{sect:dis_scattering}

\begin{figure}
\includegraphics[width=8.8cm]{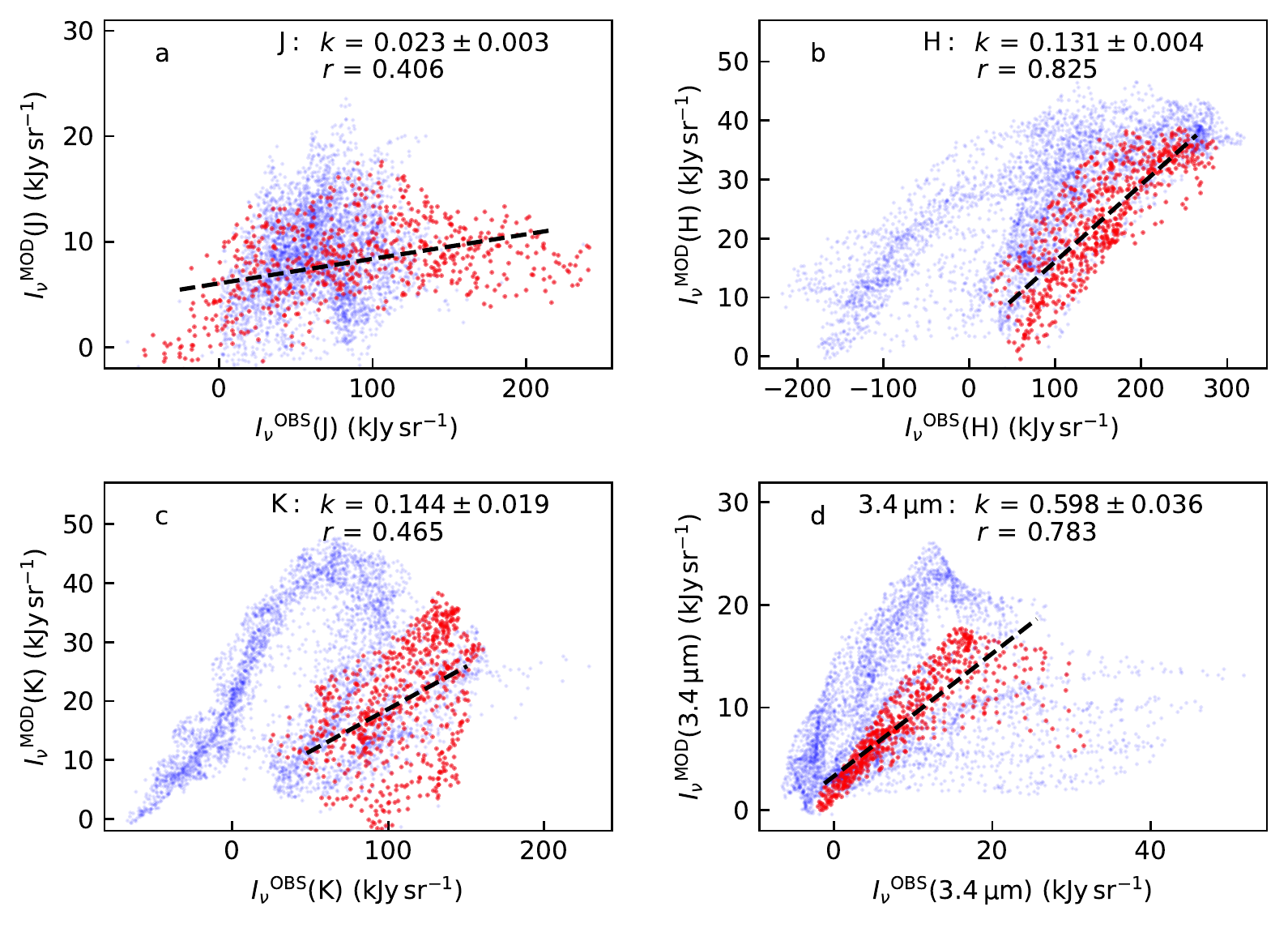}
%
%\sidecaption
\caption{
Correlation between the observed and modelled scattered light in the $J$, $H$,
$K_S$, and 3.4\,$\mu$m bands. The model results correspond to models where the
densities have been scaled down to match the NICER extinction measurements (J13
dust, and $W=1$). The data have a resolution of 25$\arcsec$ resolution and sample
at 7$\arcsec$ steps the areas marked in Fig.~\ref{fig:cmp_JHK_model_all}f. The
blue points correspond to the larger area and the red points to the smaller
area. The dashed lines show robust least squares fits to the red points. The
values of the slopes and correlation coefficients $r$ are given in the frames. 
}
\label{fig:cmp_JHK_model_corr_rescaled}
\end{figure}

\subsection{Scattering around embedded sources B2 and B3}
\label{sect:PS_models}

The large-scale model of Sect.~\ref{sect:extended} does not have the resolution
to accurately describe scattering near the embedded sources. Therefore, we made
separate, spherically symmetric models for the sources B2 and B3. The source B1
was not covered by $H$ and $K_S$ observations and is not considered here. For B2,
the observations are partly saturated but only within the innermost couple
of arcsec. 

The spherical models have a spatial resolution of 0.5$\arcsec$ (70\,au at the
distance of 140\,pc) and extend to a distance of one arcmin.
Section~\ref{sect:extended} showed that the surface brightness produced by the
external radiation field is relatively uniform at this scale. Therefore, we
computed predictions for the scattered light from the spherically symmetric
models ignoring the external illumination. When the models are compared to
observations, we subtract from both the average signal at 55-60$\arcsec$ radial
distances. The comparison will thus not be affected by the external radiation
field if its contribution is significant only at distances larger than $\sim
1\arcmin$ or if it can be approximated as a flat background. The effects of the
LOS sky background are also not considered. First, the intensity of the scattered
light within the 1$\arcmin$ region is high compared to the sky background.
Second, the effects of the $I_{\rm bg}$ are decreased by a factor of $e^{-\tau}$,
where $\tau$ refers to other LOS extinction, if that is uncorrelated with the
structure inside the 1$\arcmin$ region.

The models were fitted to observations by modifying the radial density
profile and the source luminosity. For the radial density profiles, we
tested both truncated power laws and Plummer-like functions, with no
significant difference in the fit quality. We show results for the
Plummer functions with three free parameters, i.e. the centre density
$n_0$, the characteristic radius $R$, and the asymptotic powerlaw
exponent $\alpha$,
\begin{equation}
n(r) = \frac{n_0}{[1+(r/R)^\alpha]^2}.
\end{equation}
The point sources were modelled as 6000\,K black bodies, with the
total luminosity as a free parameter. Since the scattered light is
linearly proportional to the intensity of the light sources, the
results can be easily rescaled for any assumption of the spectrum of
the central source.

\begin{figure}
\includegraphics[width=8.8cm]{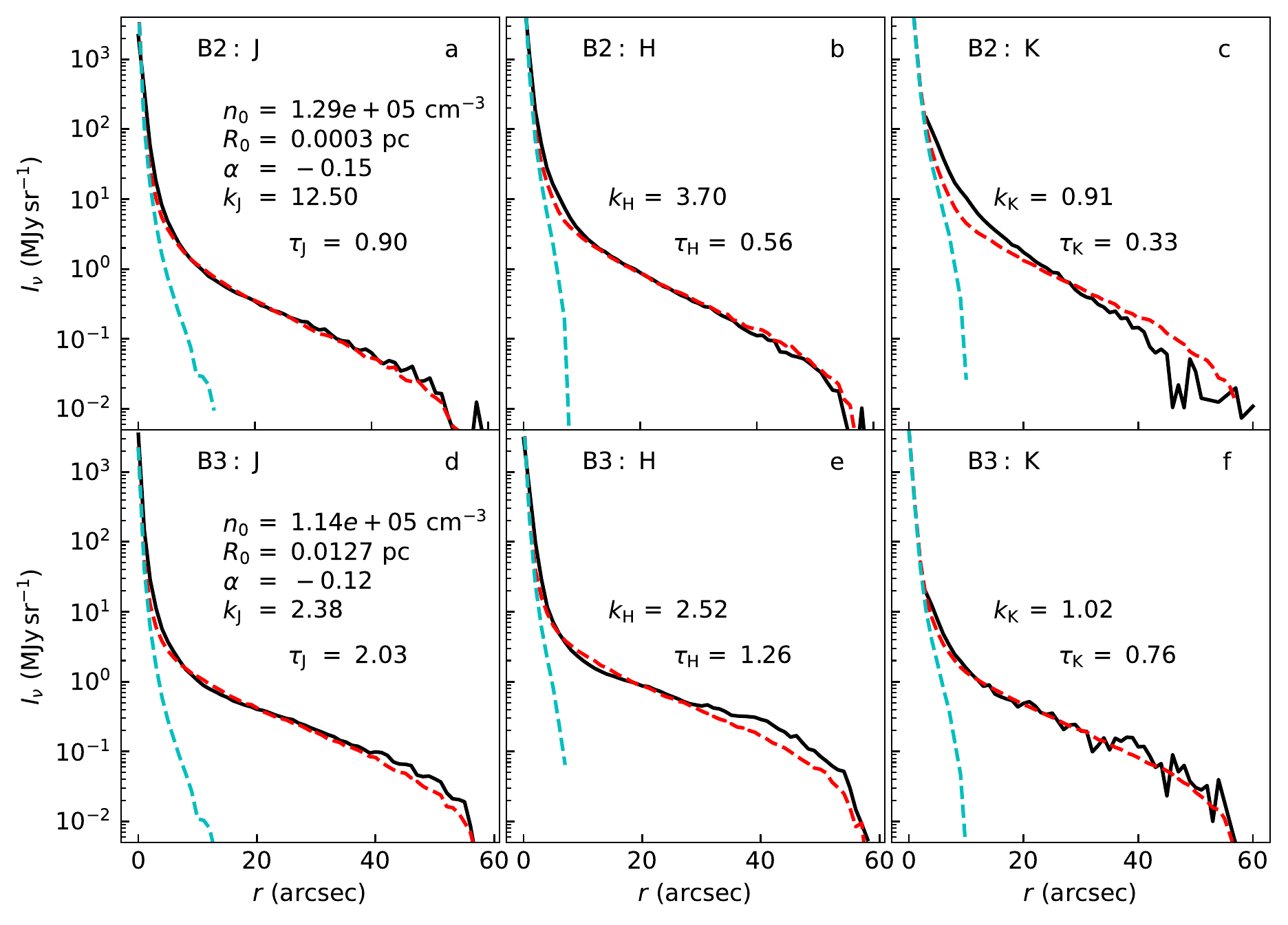}
%
%\sidecaption
\caption{
NIR surface brightness profiles in the vicinity of the embedded
sources B2 (frames a-c) and B3 (frames d-f). The black curves show the
azimuthally averaged observed profile. The red dashed curves show the
model predictions for J13 dust, consisting of the scattered light and
the attenuated direct contribution of the central source. The average
signal at 55-60$\arcsec$ distance is subtracted from both curves. The
cyan lines show the estimated direct contribution of the point source,
based on the fitted source luminosity, the LOS extinction, and the psf
shape.
}
\label{fig:plot_sphere}
\end{figure}

Figure~\ref{fig:plot_sphere} shows the results for the B2 and B3
regions (model J13), including the attenuated direct radiation from
the sources. Model maps are convolved with the point spread function
(psf) estimated from the observations of unsaturated stars in the
field. The model parameters are fitted using data at 5-40$\arcsec$
radial distances.

The figures show a good match between the observations and the model,
which in the 10-40$\arcsec$ distance range mainly consists of
scattered light. Towards the centre, the signal is dominated by the
direct light from the point source and unresolved scattering. Also
this part is well matched, the intensity profiles following the shape
of the point spread function. The psf was estimated up to a radial
distance of 15$\arcsec$ but there it is already orders of magnitude below
the peak value and cannot be measured reliably. In the outer part,
beyond 40$\arcsec$, the final drop is caused by the background
subtraction. 

The best fit density profiles are practically flat with $\alpha
\ga-0.15$ and the observed surface brightness is almost consistent
with a radial decrease of the radiation field intensity in a 
homogeneous medium. The LOS optical depths from the source to the
observer are for B2 below one and for B3 of the order of one. The
contribution of thermal emission should be small for the assumed
6000\,K sources \citep{Sellgren1996}, because of the low colour
temperature and because of the large extinction that further removes
short-wavelength photons from the radiation field. The fitted
luminosities (see Fig.~\ref{fig:plot_sphere}) suggest higher
effective temperatures for the central sources, but these values
should be sensitive to the assumed structure and opacity of the dust
layers closest to the central source. These are not well constrained
by the available data. Because the relative contribution of thermal
emission should decrease with radial distance (as short-wavelength
photons are absorbed), it should result in steeper density profiles
in the models where only the scattered light is considered.
Therefore, the fact that the derived density profiles were flat
argues against any significant contribution from dust emission.

Appendix~\ref{app:sca} shows corresponding plots for models with AMMI
dust. Because of the higher dust albedo, the derived source
luminosities are lower by almost a factor of two but the derived
density profiles are still flat. For B2, the powerlaw exponent is
smaller, $\alpha=-0.97$, but the characteristic radius is large,
$R_0=0.016$\,pc, which corresponds to $\sim26\arcsec$. Thus, although
the fit parameters are partly degenerate, all models show only little
radial density variation. The degeneracy also applies to the
parameters $n_0$ and $R_0$.

\section{Discussion}  \label{sect:discussion}

\subsection{Sub-millimetre dust opacity} \label{dis:opacity}

We derived with HAWK-I NIR observations and Herschel sub-millimetre
data three estimates for the submm-to-NIR extinction ratio that were
consistent with $\tau(250 \, \mu {\rm m})/ \tau(J) = (1.0 \pm
0.2)\times 10^{-3}$.
The comparison of data at 2$\arcmin$ resolution gave $\langle \tau(250 \, \mu
{\rm m})\rangle / \langle \tau(J)\rangle = 1.07\times 10^{-3}$ while the slope of
the least-squares fit gave $\tau(250\,\mu {\rm m})/ \tau(J) = (1.22 \pm
0.04)\times 10^{-3}$. The latter fit excluded low-column-density regions and the
higher value could thus be more representative of the central regions of
LDN~1642. As discussed in \citet{GCC-V}, both optical depth estimates
$\tau(250\,\mu{\rm m})$ and $\tau(J)$ could be biased. 
%For extinction, this could be
%associated with steep optical depth gradients and lower stellar
%densities. When the fit of $\tau(250\,\mu{\rm m})$ and $\tau(J)$ was
%repeated using individual stars, the obtained value was indeed smaller
%but not significantly so, $\tau(250\,\mu {\rm m})/ \tau(J) = (0.95 \pm
%0.04)\times 10^{-3}$. The small-scale column density structure was
%already partly taken into account in the making of the extinction maps
%(Sect.~\ref{sect:colden}). 
However, because of the modest optical depths, the bias in
$\tau(250\,\mu{\rm m})$ should remain small. The maps of $\tau(250\,\mu{\rm
m})/\tau(J)$ were quite flat, also suggesting that the ratio could be measured
reliably. 

LDN~1642 was in the sample of cold clumps analysed in \citet{GCC-V} as
source G210.90-36.55, where the combination of Herschel and 2MASS data resulted
in an higher value of $\tau(250\,\mu {\rm m})/ \tau(J) = 1.6\times 10^{-3}$. This
value was close to the median ratio for a sample of clouds extracted from the
Planck survey of cold clumps \citep{planck2014-a37}. The sub-millimetre opacity
was obtained from MBB fits with $\beta=2.0$ and with $\beta=1.8$ (as used in the
present paper), the value would decrease by some 20-30\%, becoming marginally
consistent with our new results. 
%In \citet{GCC-V},
%the direct $\tau(250\,\mu{\rm m})$ and $\tau(J)$ estimates were corrected for the
%expected bias with the help of modelling. On the other hand, that comparison
%relied on 2MASS data and, because of the smaller number of stars, the analysis
%was done at three arcmin resolution.

The ratio $\tau(250\, \mu {\rm m})/ \tau(J)$ of LDN~1642 is significantly higher
than in the diffuse medium. Assuming the $R_{\rm V}=3.1$ extinction curve and the
ratio $N({\rm H}_2)/A_V = 9.4 \times 10^{20}$\,cm$^{-2}$\,mag$^{-1}$ \citep{BSD}
our result corresponds to $\tau(250\,\mu{\rm m})/N_{\rm H}=1.5 \times
10^{-25}$\,cm$^2$\,H$^{-1}$. This value is three times higher than
the Planck measurement $\tau(250\,\mu{\rm m}/N_{\rm H}) \sim 0.5 \times
10^{-25}$\,cm$^2$\,H$^{-1}$ obtained at high latitudes. We scaled the value to
250\,$\mu$m using the opacity scaling $\nu^{1.53}$ given in that paper
\citep{planck2013-XVII}. \citet{Fukui2014} examined atomic regions around
selected high-latitude clouds and obtained a value $\tau(850\,\mu{\rm m})/N_{\rm
H}= 1.5 \times 10^{-26}$\,cm$^2$\,H$^{-1}$. For spectral indices $\beta=1.5-1.8$,
this results in similarly low 250\,$\mu$m dust opacities,
$\tau(250\,\mu{\rm m}/N_{\rm H}) = (0.4-0.6) \times
10^{-25}$\,cm$^2$\,H$^{-1}$. 

The LDN~1642 value is similar to previous measurements of dense molecular clouds,
indicating clear dust evolution relative to the diffuse medium
\citep{Stepnik2003,Martin2012,Roy2013}. Below we mention some recent studies.
When the original results were reported for different wavelengths, we assume
$\beta=1.8$ for the long wavelengths and the standard extinction curve for NIR
($R_V=3.1$), to scale the results to $\tau(250\,\mu {\rm m})/ \tau(J)$.

\citet{Suutarinen2013} used Herschel data and dedicated NIR observations to
derive directly a value $\tau(250\,\mu {\rm m})/ \tau(J) =1.4\times 10^{-4}$ for
the Corona Australis cloud. \citet{Lombardi2014} used both Planck and Herschel
data to derive ratios of 850\,$\mu$m and $K_S$-band opacity in Orion. The results
correspond to $\tau(250\,\mu {\rm m})/ \tau(J)=1.5\times 10^{-3}$ and $1.1\times
10^{-3}$ for the Orion clouds A and B, respectively. With similar analysis,
\citet{Zari2016} found a value of $1.0\times 10^{-3}$ in the Perseus molecular
cloud and \citet{Lada2017} found $1.1\times 10^{-3}$ in the California
Nebula. Even larger relative increases of dust opacity have been reported at
longer wavelengths \citep{Mason2020}.
On the other hand, \citet{Forbrich2015} found for FeSt~1-457 (a core in the Pipe
nebula) a dust opacity that was only slightly higher than in the diffuse medium,
$\tau(250\,\mu {\rm m})/ \tau(J)=0.65\times 10^{-3}$, in spite of the data
covering extinctions up to $A_{\rm K}=5$\,mag. 

The high $\tau(250\,\mu {\rm m})/ \tau(J)$ values are more consistent with models
of evolved dust, with increased grain sizes and the formation of aggregates,
possibly covered by ice \citep{OH94,Ormel2011,Kohler2012,Ysard2016, Ysard2019}.
Corresponding changes should exist also in the shorter-wavelength scattering
properties of the grains.

\subsection{Scattered light}  \label{dis:scattering}

The light scattered by dust grains has been observed in the NIR, referred as
cloudshine, towards many molecular clouds (\citealt{2008A&A...480..445J},
\citealt{2012A&A...544A..14J}, \citealt{Lefevre2014}). However, recently
discovered MIR dust scattering through Spitzer 3.6 $\mu$m and WISE 3.4\,$\mu$m
data, referred as coreshine, was the first direct indication of a significant
population of $\sim$1 $\mu$m grains in pre-stellar cores (\citealt{Pagani2010},
\citealt{Steinacker2010}, \citealt{2012A&A...544A..14J}).
\cite{2008A&A...480..445J, 2009A&A...505..663J, 2012A&A...544A..14J} studied the
Corona Australis cloud in detail and concluded that NIR scattered light can be
used to estimate better resolution column density maps at low visual extinction
($A_V \la 10$). They also found a linear relationship between Herschel
column density estimates and NIR scattering.

Our data included HAWK-I NIR and WISE 3.4\,$\mu$m measurements. We have assumed
that the surface brightness is due to scattered light only.
Figure~\ref{fig:surfb} clearly shows extended emission from the densest part of
the LDN 1642 cloud. \cite{Malinen2014} studied the LDN~1642 cloud at multiple
wavelengths and suggested that the extended emission from the densest part of LDN
1642 in the WISE 3.4\,$\mu$m image could be due to scattered MIR light (the
coreshine phenomenon) associated with grain growth \citep{Steinacker2010}.
% However, they also suggested possible contributions from the psf tails of the
% three bright YSOs and additional scattering from the embedded sources. 
We masked the bright sources in the WISE 3.4\,$\mu$m image and showed that the
extended emission is similar to the NIR surface brightness and the column density
maps. It can be considered an upper limit for dust scattering. No extended
emission is found in the WISE 4.5 $\mu$m map. In the comparison with the Herschel
column density, we found a strong linear correlation in the $J$, $H$, and
3.4\,$\mu$m bands.  The correlation was weaker in the $K_S$ band because of
the smaller optical depth and problems with the data quality.

\cite{Lefevre2014} studied dust grain properties inside molecular clouds using coreshine modelling and
showed how the intensity of the coreshine depends on the incident radiation, the extinction of the
background radiation, the grain properties, and the core properties. They found a higher NIR/MIR (K/3.6
$\mu$m) ratio for the Taurus-Perseus region, which could be explained by the presence of ice mantles.
Alternatively, the grain size distribution having bigger silicates than carbonaceous grains could
explain the higher NIR/MIR ratio \citep{Lefevre2014}. For LDN~1642, the NIR/MIR ($K$/3.6 $\mu$m) ratio
is $\sim$8 (11) for the larger (smaller) region shown in Fig.~\ref{fig:cmp_JHK_model_all}f. These high
values are consistent with that of the Taurus-Perseus region, in that interpretation suggesting bigger
silicate dust grains \citep{Lefevre2014}. 
%% Much higher values could also be due to the presence of the three YSOs in the cloud. 
The $J$/$K$ ratio for LDN~1642 is $\sim$1.6 (0.9) for the larger (smaller) region.
These values are also similar to the typically observed range of 0.3-3
found by \cite{Lefevre2014}.

\subsection{Radiative transfer models}  \label{dis:RT}

In this section, we discuss the interpretation of the results based on
the RT modelling. We start with models fitted to the sub-millimetre
emission and their consistency with NIR extinction. We then discuss
the NIR surface brightness and the tension between the models matching
observations of either the sub-millimetre dust emission or the
combination of NIR scattering and extinction.

\subsubsection{Models of dust emission} \label{dis:RT_emission}

%RT models were fitted to the 250-500\,$\mu$m sub-millimetre emission,
%to characterise the cloud structure and partly the effects of the dust
%emission properties. The obtained density fields also provided the
%starting point for the subsequent NIR scattering calculations.

The fit residuals were mostly below 10\% for all sub-millimetre bands and
comparable to the observational uncertainties (Fig.~\ref{fig:mod_emit}) and even
the 160\,$\mu$m extrapolated values were almost at the correct level. The fit
quality was mostly independent of the tested dust models
(Fig.~\ref{fig:mod_emit_2}) and large residuals are not expected because of the
large number of free parameters. However, because the radiation field scaling
affects the whole map, differences in dust opacities (sub-millimetre vs. optical)
could have caused systematic differences in the relative colour temperature of
between dense and diffuse regions. There tended to be positive residuals at short
wavelengths, partly correlated with the column density. These residuals
suggest that the dust temperature is too low in the dense part of the cloud. In
agreement with this possibility, the residuals were smaller in the $W=1.7$
case when the cloud structure allowed more radiation to reach the cloud centre.
The residuals could also reflect dust property variations that reduce the
short-wavelength cloud opacity. However, the sub-millimetre data alone do not
give strong constraints on dust models. 

\begin{figure*}
\includegraphics[width=18cm]{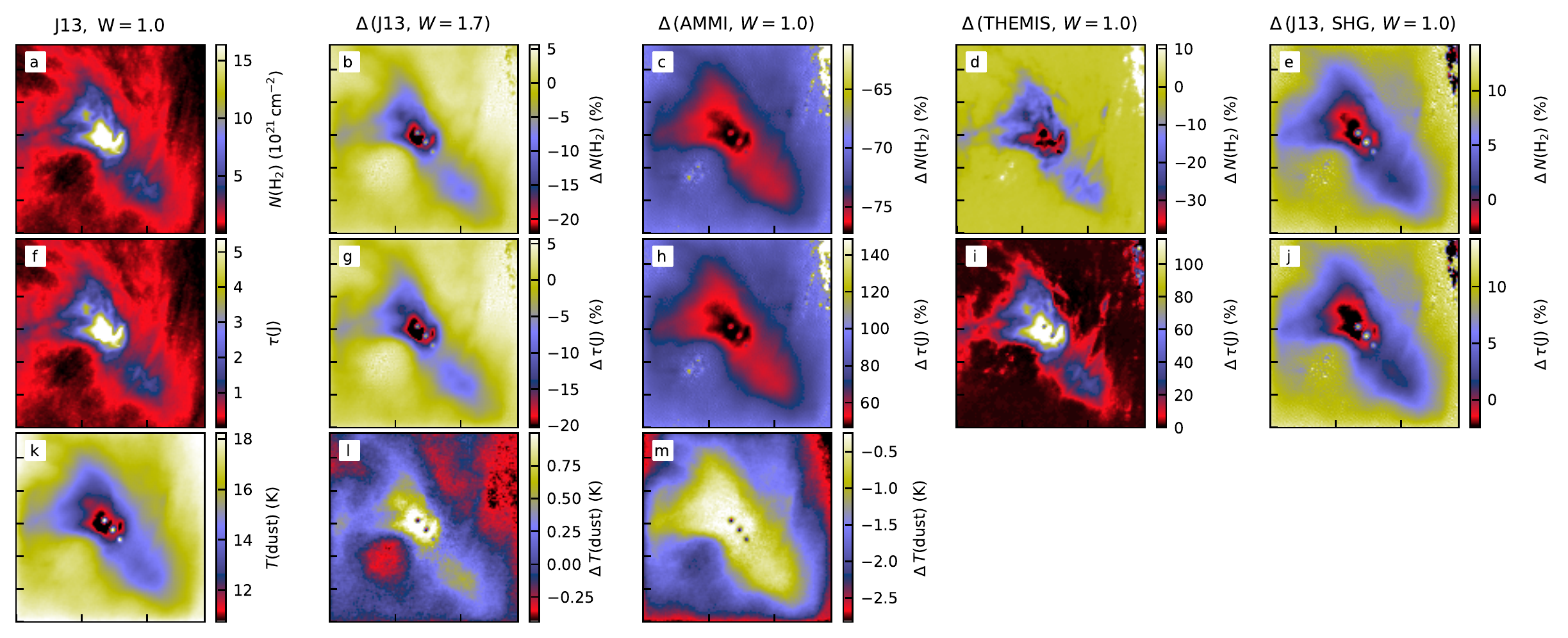}
%
%\sidecaption
\caption{
Comparison of model column density $N({\rm H}_2)$ (first row), NIR
optical depth $\tau(J)$ (second row), and a cross section of the
physical dust temperature $T({\rm dust})$ (third row) in alternative
models. The first column shows the reference model with J13 dust and
$W=1$ LOS cloud extent. The other frames show differences to this
model, for the dust and $W$ values indicated on top. The last column
(J13, SHG) is the same as the reference model but with calculations
including the full treatment of stochastic heating. Temperature maps
are not shown for the THEMIS and stochastically-heated-grain cases,
where there is no single temperature per cell.
}
\label{fig:plot_ntt_shg}
\end{figure*}

Although the sub-millimetre data can be fitted with any of the dust models, they
do lead to significant differences in other parameters.
Figure~\ref{fig:plot_ntt_shg} shows the changes in the column density, $J$-band
optical depth, and dust temperature relative to the J13 model. 

%% TEMPERATURE
In the case of AMMI, the central temperature is almost unchanged while in the
outer parts there is an over 2\,K drop relative to the J13 case. Although
ice-coated grains are not be expected in diffuse regions, they can affect the
temperature profiles deeper in the cloud. The dust temperature is an important
parameter for cloud and core chemistry and even for their gravitational stability
\citep{Bergin2006,Sipila2017}. In these respects, even differences of 1\,K can be
significant.

%% COLUMN DENSITY
The dust differences also affect the estimated mass distributions.
Assumption of higher sub-millimetre emissivity would result in lower optical
depths in the UV-optical-NIR regime, thus leading to a more uniform temperature
distribution and smaller column-density variations. Such muted differences
would clearly be noted if the comparison was made between models with the
same radiation field. In our case, these effects are partly compensated by
changes in the ISRF level. For AMMI dust, the column density is up to 75\% lower
than for the J13 model. The reduction is some 10\% larger towards the cloud
centre than in the outer cloud regions. Such trends are significant for the
accuracy of mass estimates but also for estimates of the core density profiles.
The spatial dust-property variations in real clouds naturally introduces
additional uncertainties. 
%In the
%THEMIS model, the reduction of $N({\rm H}_2)$ is on average smaller than in the
%pure AMMI case, but there are locally steeper variations associated with the
%changes in the relative abundance of the dust components.

\begin{figure}
\includegraphics[width=8.8cm]{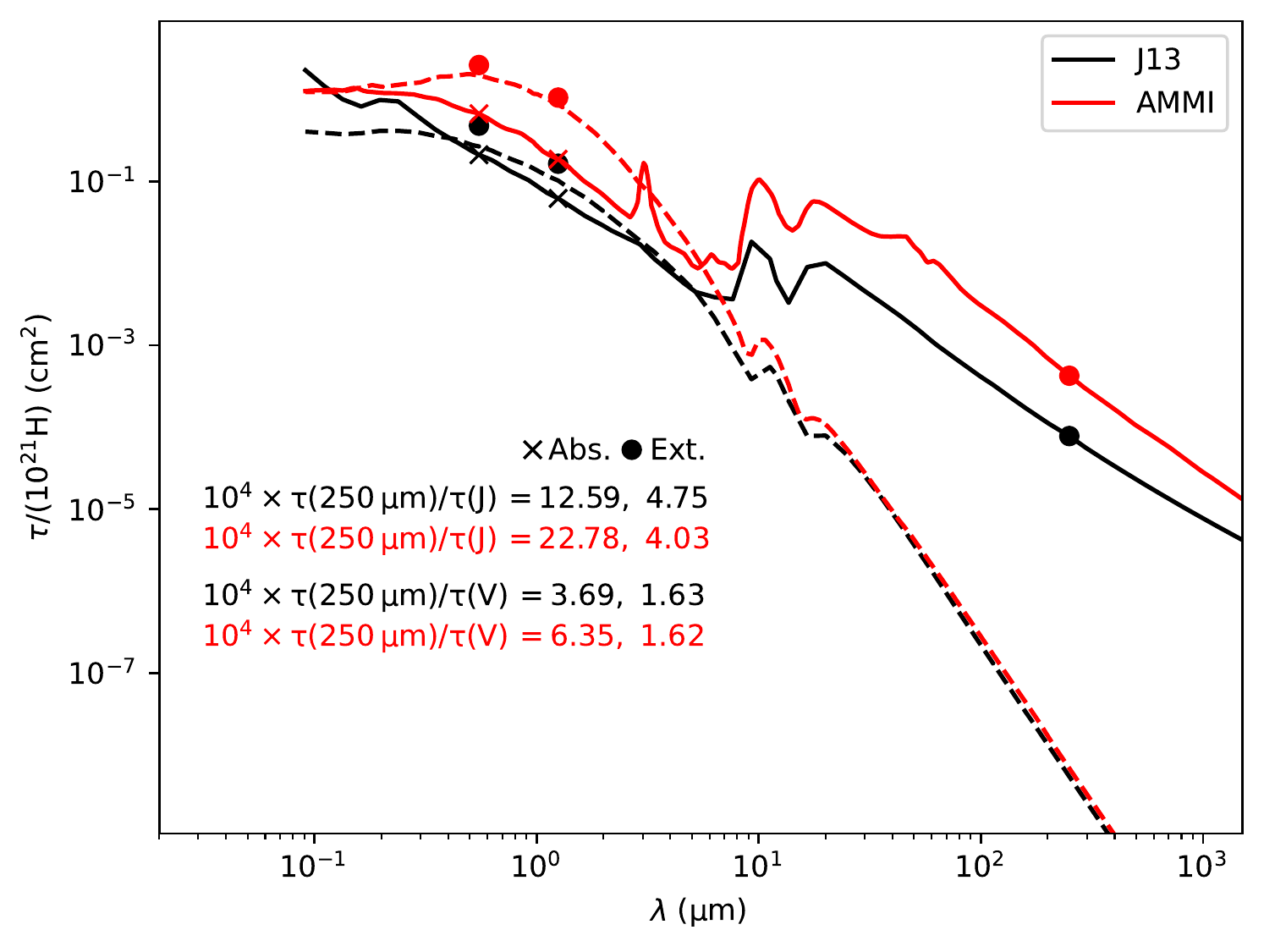}
%
%\sidecaption
\caption{
Dust optical depth for J13 and AMMI models relative to hydrogen column density
$N({\rm H})=10^{21}\,{\rm cm}^{-2}$. The solid lines are for absorption and the
dashed lines for scattering. The plot lists $\tau(250\,\mu{\rm m})/\tau(J)$
ratios where the first number corresponds to absorption (points marked with
crosses) and the second to extinction (marked with circles).
}
\label{fig:tau_curves}
\end{figure}

%  NIR OPTICAL DEPTH
In spite of the lower model column densities, AMMI and THEMIS led to 40-100\%
increases in the predicted NIR opacity because both dusts have significantly
higher opacities also at NIR wavelengths. This is illustrated by
Fig.~\ref{fig:tau_curves} that compares the J13 and AMMI opacity curves.
Although the average increase in $\tau(J)$ is similar for AMMI and THEMIS
(Fig.~\ref{fig:plot_ntt_shg} and Table~\ref{table:tau_J}), the spatial
distributions of the $\tau({\rm J})$ increase are different. For AMMI the
increase is smallest towards the cloud centre while for THEMIS it follows the
dust abundance variations and peaks at the cloud centre. Such differences
will be reflected in the contrast of the scattered-light images.

Figure~\ref{fig:plot_ntt_shg} highlights two further factors affecting the
emission modelling. If the cloud is more extended in the LOS direction, the
optical depths become lower in the perpendicular direction, the central
temperature rises, and the observed surface brightness is reproduced with a lower
column density. When the LOS extent was increased by 70\% ($W=1.7$), the column
density was up to $\sim$20\% lower and the core temperature higher by $\sim$1\,K.
These effects are thus almost of similar magnitude as the differences between the
dust models.

Because of the higher computational cost of the full treatment of stochastically
heated grains, the long-wavelength emission was calculated assuming grains at an
equilibrium temperature, which is true only for large grains. The absorption and
MIR emission by smaller grains is energetically significant and when part of the
emitted energy is transferred to shorter wavelengths, a higher column density is
needed to produce a given sub-millimetre intensity. Figure~\ref{fig:plot_ntt_shg}
includes results for the J13 model ($W=$1.0) when the full grain size
distributions and the stochastic heating are taken into account. The column
densities of the fitted model are higher by up to $\sim$10\% in the outer cloud
regions, where the MIR emission is strong. The difference decreases with column
density and is only a couple of per cent towards the cloud centre. For the most
accurate results, the full treatment of grain heating would still be preferred,
if computationally feasible. For our models consisting of $\sim 4\times 10^6$
cells, the full treatment slowed down the calculations by more than a
factor of ten, to about one hour per iteration.

We assumed the \citet{Mathis1983} ISRF model as the reference and included the
scaling $k_{\rm ISRF}$ as a free parameter. The spectral shape of the incoming
radiation has additional second-order effects. The cloud models cover a limited
volume, which can be assumed to be surrounded by outer cloud layers that
attenuate the incoming radiation.  By preferentially removing short-wavelength
radiation, an external layer would increase the mean free path for the remaining
radiation, resulting in smaller temperature gradients. The observations used in
the modelling were similarly background-subtracted, to eliminate the extended
foreground and background emission. Using these original surface brightness data
with absolute zero points (Sect.~\ref{sect:Herschel_observations}) and the J13
dust properties, we estimate $\tau(J)=$0.52 for the average optical depth in the
area used for background subtraction. If this corresponds to a layer between
LDN~1642 (the modelled volume) and the stars providing the ISRF, the incoming
radiation should be attenuated by $e^{-\tau_{\rm Ext}(\nu)}$, where the optical
depth of the external layer could be up to $\tau(J)\approx$0.26, half of the LOS
value. We recomputed the J13 model with this change in the shape of the incoming
ISRF radiation. The optical depth of the optimised model decreased only by 5\%.
The effect of $\tau_{\rm Ext}$ is likely to be even smaller, since at least part
of the LOS material is mixed with the stellar distribution. Therefore, the
discrepancy in the $\tau(J)$ values (models vs. NICER) can not be resolved by an
external cloud layer or other similar changes in the shape of the ISRF spectrum.

\subsubsection{Models of scattered light} \label{sect:dis_scattering}

The morphology of the model predictions was consistent with the observations of
the large-scale scattering but only if the background component $I_{\rm
bg}(e^{-\tau}-1)$ was ignored. Better results were obtained by using model clouds
with lower column density, in agreement with the direct NIR extinction
measurements.

\begin{figure}
\includegraphics[width=8.5cm]{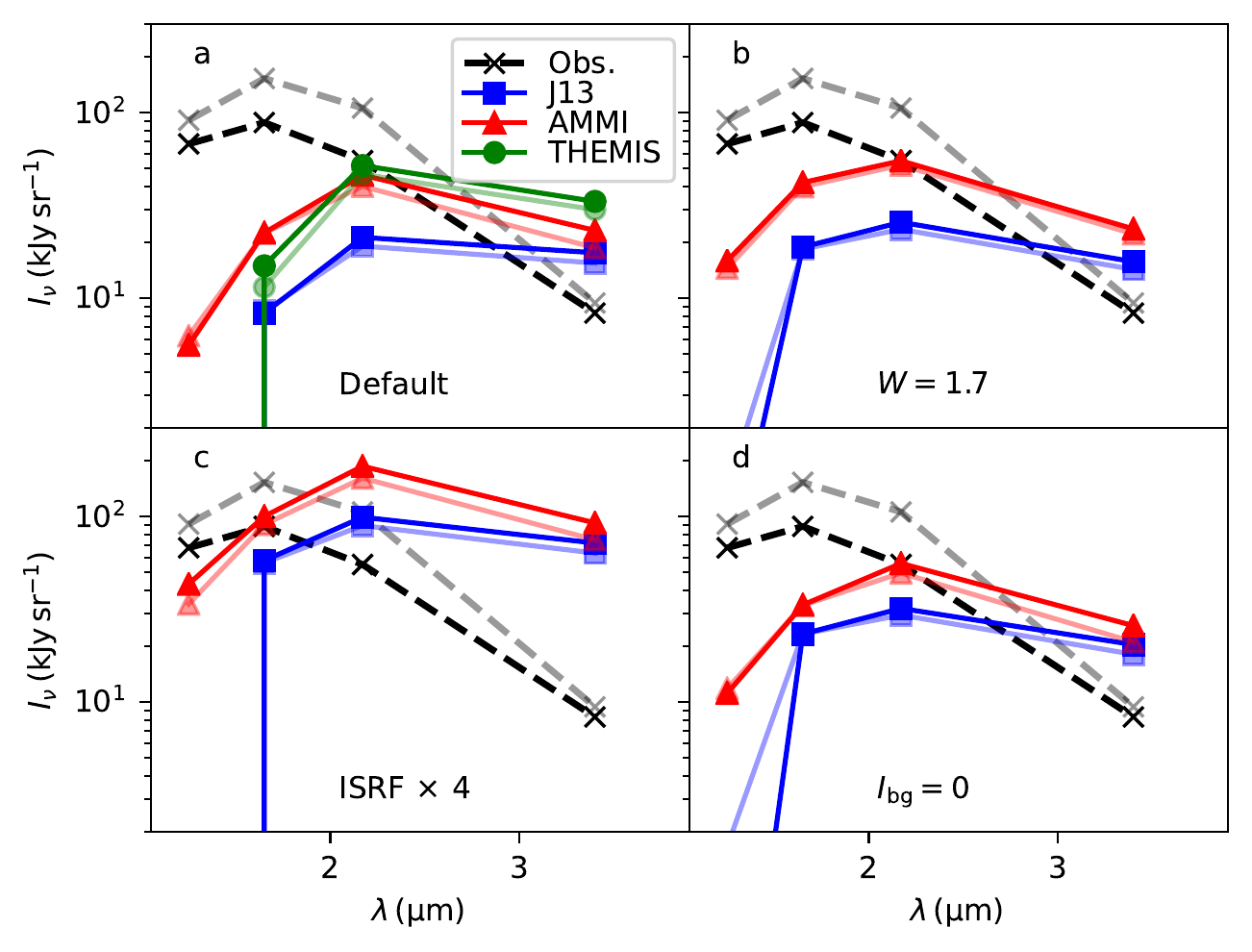}
%
%\sidecaption
\caption{
Comparison of observed net surface brightness (dashed lines) and
models with J13, AMMI, and THEMIS cases. The frames correspond to
different model assumptions: the default parameters (frame a), cloud
model with larger LOS extent ($W=1.7$, frame b), four times larger
intensity of the local radiation field (frame c), and the default
model assuming no LOS sky background ($I_{\rm bg}$=0, frame d).
Estimates
Fig.~\ref{fig:cmp_JHK_model_all} and include estimates for the areas
marked in Fig.~\ref{fig:cmp_JHK_model_all}. The solid lines correspond
to the larger area and the partly transparent lines to the smaller
area. THEMIS results are shown only for the default case.
}
\label{fig:plot_JHK_vs_alternatives}
\end{figure}

When we used model clouds obtained from the emission modelling, the NIR
observations were underestimated most severely at the shortest wavelengths. The
$J$-band surface brightness was mostly negative and the $K_S$-band and
3.4\,$\mu$m signals were half those observed (Fig.\ref{fig:cmp_JHK_model_all}).
Similar wavelength dependences existed for all dust models. This behaviour
was suggestive of excessive NIR optical depths in the cloud models.
In addition to the column density, the surface brightness depends on the dust
model, the 3D shape of the cloud, and the sky brightness behind the cloud. We
summarise these effects in Fig.~\ref{fig:plot_JHK_vs_alternatives}, for the cloud
models obtained from the fitting of dust emission.

AMMI dust increases the levels of scattered light but while the observed signal
is underestimated at the shortest wavelengths, the 3.4\,$\mu$m prediction is too
high. Apart from the level of the NIR signal, the change of the dust model does
not improve the match to the observed shape of the SED. The larger LOS cloud
extent clearly increases the predicted surface brightness and changes the SED
shape in the correct direction. However, the effect remains too small.

The NIR modelling adopted the \citet{Mathis1983} ISRF values but, because the
scattered light is directly proportional to the incoming radiation, the results
could be easily rescaled.  For example, the DIRBE measurements suggest some 40\%
higher NIR intensities \citep{Lehtinen1996}. As shown in
Fig.~\ref{fig:plot_JHK_vs_alternatives}c, we would need a much larger factor to
match the general level of the NIR observations and, if the same scaling was
applied to all bands, the SED shape would still not match the observations. The
1.25-3.4\,$\mu$m radiation field had an effective colour temperature of 3500\,K.
Even if the field were dominated by 10000\,K sources, the $J$-band to 3.4\,$\mu$m
ratio would increase by less than a factor of three. In the absence of
nearby massive stars, the discrepancy in the level and spectral shape of the NIR
excess cannot be solved by modifications of the radiation field either.

If sky brightness behind the cloud were severely overestimated and we set $I_{\rm
bg}$ to zero (Fig.~\ref{fig:plot_JHK_vs_alternatives}d), the NIR signal would 
increase but this would still not fix the incorrect SED shape. If the ISRF is
boosted to increase the scattering, the background term becomes relatively small
and has only little effect on the NIR spectrum. Therefore, none of the above
modifications can resolve the discrepancy between the observed and modelled NIR
surface brightness. Of course, the independent NIR extinction measurements
already suggested that the main problem resides in the high NIR optical depth.

A higher optical depth of an already optically thick cloud can directly decrease
the intensity of the scattered light \citep{Juvela2006_sca}, at the same time
making the term $I_{\rm bg}(e^{-\tau}-1)$ more negative.
Section~\ref{dis:RT_emission} noted that the discrepancy between the NIR optical
depths deduced from the dust emission models and the direct extinction
measurements was between a factor of 2.3 and 4.6 (Table~\ref{table:tau_J}). When
the calculations were repeated with cloud models with $\tau(J)$ decreased to
match the NICER estimates, the results were clearly improved
(Fig.~\ref{fig:cmp_JHK_model_corr_rescaled}).

\begin{figure}
\includegraphics[width=8.5cm]{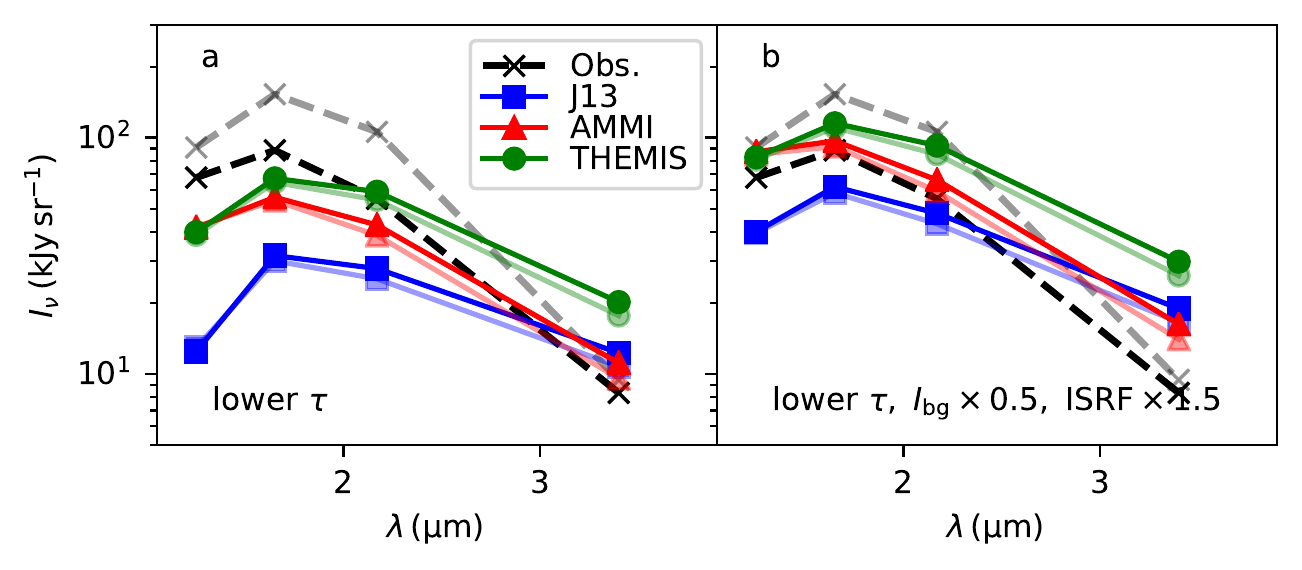}
%
%\sidecaption
\caption{
Comparison of observed net surface brightness (dashed lines) and model
predictions for J13 and AMMI dust. The density of the cloud models
has been rescaled to match the NIR extinction measurements. In frame
b, the general intensity of the ISRF has been scaled further by a
factor of 1.5 and the intensity of the sky background $I_{\rm bg}$ by
0.5.
}
\label{fig:plot_JHK_vs_alternatives_2}
\end{figure}

Figure~\ref{fig:plot_JHK_vs_alternatives_2}a summarises the results for
lower-density cloud models when other parameters ($I_{\rm}$, $W$, and ISRF) are
kept at their default values. The predicted spectra are now closer to the
observations. The J13 model remains in the J band a factor of several below the
observations. For the AMMI model both the intensity level and SED shape are much
closer to the observations. THEMIS provides a higher average NIR brightness but
its SED shape is more inconsistent with the observations. We emphasise that
Fig.~\ref{fig:plot_JHK_vs_alternatives_2} shows the surface brightness excess
relative to the reference area shown in Fig.~\ref{fig:cmp_JHK_model_all}.
Therefore, the THEMIS results are sensitive to the density thresholds used in
setting the relative abundances of the dust components. The reference area is
mainly below the densities where the final transition to AMMI dust takes place.
If this transition took place at a lower density, the THEMIS result would become
more similar to the AMMI one.

In Fig.~\ref{fig:plot_JHK_vs_alternatives_2}b the ISRF is further assumed to be
50\% higher and the LOS background is 50\% lower. The first change would be in
agreement with the DIRBE measurements, as discussed above. A 50\% error in the
$I_{\rm bg}$ is unlikely but the effect of $I_{\rm bg}$ is already relatively
small, because of the lower optical depth and the higher intensity of the
scattered light. With these changes, the J13 predictions increase significantly
but remain below the observed values at the shortest wavelengths. AMMI matches
the NIR data well, while the 3.4\,$\mu$m value is slightly overestimated. For
THEMIS, the MIR signal is clearly above the observations.

A further reduction in model column density should improve the match with the
observed SED shape, increasing the short-wavelength signal relative to the longer
wavelengths. However, the SED shape also depends on other factors, such as the
details of the spatial variations of dust properties. We conclude that the
LDN~1642 NIR observations can be explained by using models with evolved dust
components, such as AMMI. The types of dust grains found in the diffuse ISM are
not able to provide sufficient surface brightness or the correct spectral shape,
not without improbable modifications to several parameters, including the
intensity and the spectral shape of the external radiation field.

The modelling of the immediate environment of the sources B2 and B3
showed that the surface brightness is compatible with pure scattering
(Sect.~\ref{sect:PS_models}). The observations should have been
sensitive to the potential additional NIR component of dust emission,
unless that is restricted to within 10$\arcsec$ ($\sim0.01$\,pc) of
the sources. The similarity of the surface brightness profiles between
B2 and B3, and the good match with the scattered-light models also
confirm that B3 is indeed part of the LDN~1642 cloud.

\subsubsection{Discrepancy between emission and scattering}
\label{dis:discrepancy}

A major discrepancy existed between the higher NIR optical depths predicted by
the dust emission modelling and, on the other hand, the lower values suggested
both by the simulations of the NIR scattering and the direct extinction
measurements with background stars. If the NIR optical depth could be reduced,
the NIR surface brightness could be explained with dust with high albedos, such
as AMMI.

The $W=1$ and $W=1.7$ models showed that the cloud shape does not provide a
solution, possibly with the exception of the unlikely scenario of a very long
filament viewed along the main axis. As an alternative to limit the dust
temperature variations (and thus to decrease the column density), we also briefly
tested the effects of a clumpy cloud structure by scaling the density values with
Gaussian random fields with different powerlaw indices, with $\sigma=1$ on
logarithmic scale or with direct multiplication with $N(\mu=1, \sigma=0.35)$.
However, the effects on the predicted $\tau(J)$ remained smaller than the
difference between the $W=1$ and $W=1.7$ models.

Of the dust properties, the albedo and the scattering function are important for
the NIR surface brightness but do not affect the main problem of the NIR
extinction. The problem does not concern only the modelling but was already shown
by the empirical result $\tau(250\,\mu{\rm m})/\tau(J)\sim 10^{-3}$. This ratio
is a factor of two lower for J13, $\tau(250\,\mu{\rm m})/\tau_J=0.49 \times
10^{-3}$, and even lower for AMMI, $0.40 \times 10^{-3}$. 

The difference in the NIR extinction curve of the three dust models had a
negligible effect on the $\tau(J)$ values (Table~\ref{table:tau_J}). Previous
studies also have concluded the NIR extinction curve to be relatively constant,
with variations at most at a 5 percent level \citep{Indebetouw2005,
Lombardi2006_Pipe, RomanZuniga2007, SteadHoare2009, Fritz2011, Ascenso2013,
WangJiang2014}. Some of these studies have targeted clouds with column densities
higher than in LDN~1642. Our estimate $E({\rm H-K})/E({\rm J-H})=0.73\pm0.35$ was
fully consistent with the standard extinction curve, given its large uncertainty.
The uncertainty caused by the shape of the extinction curve is thus likely to be
below $\sim$10\%. Small-scale cloud structure could bias $\tau(J)$ values but in
the other direction, reducing the extinction estimates \citep{Lombardi2009}. 
Down to 40$\arcsec$ scales, the small-scale structure was already taken into
account with the help of Herschel observations. The ratio between the $\tau(J)$
values from the emission models and from the NICER calculations was spatially
constant, which also suggests that errors related to cloud gradients or
variations in the local stellar density are not significant.

Assuming that the observed $\tau(J)$ values are accurate and taking the observed
$\tau(250\,\mu{\rm m})/\tau(J)$ ratio as the starting point, we tested ad hoc
dust models where the opacities at $\lambda>60\,\mu$m were scaled with a constant
to match $\tau(250\,\mu{\rm m})/\tau(J)=10^{-3}$. When these modified dusts were
used in the emission modelling, the $\tau(J)$ values were reduced almost
proportionally to the increase of the sub-millimetre opacity
(Table~\ref{table:tau_J_adhoc}). For J13, the $\tau(J)$ value of the emission
model is nearly consistent with the NICER estimate while for AMMI there still
remains a factor of two discrepancy. With the modified dust models, the radiation
field estimates were increased to $k_{\rm ISRF}\sim$1.2-1.5, thus mainly between
the lower \citet{Mathis1983} estimates and the higher values obtained from DIRBE
observations \citep{Lehtinen1996}.
                               
\begin{table}
\caption{NIR optical depths of model clouds ($\tau_J^{\rm M}$)
relative to NICER measurements ($\tau_J^{\rm N}$) for ad hoc
dust models with larger sub-millimetre emissivity.}
\begin{tabular}{lcccc}
\hline \hline
Model    & 
$\langle \tau_J^{\rm M}  \rangle$  &  
$\langle \tau_J^{\rm N} \rangle$  &  
$\langle \tau_J^{\rm M}  \rangle$ / $\langle \tau_J^{\rm N} \rangle$  &
$k_{\rm ISRF}$ \\
\hline
  J13   &  0.19  &  0.17  &   1.15  &  1.21 \\
%  COM   &  0.21  &  0.16  &   1.25  &  1.14 \\
%  CMM   &  0.34  &  0.15  &   2.17  &  0.93 \\
  AMMI    &  0.32  &  0.15  &   2.22  &  1.48 \\
  THEMIS  &  0.20  &  0.17  &   1.17  &  1.21 \\  %% THEMIS2X  !!!
% THEMISX_W1.00  & 0.20 &  0.17  &   1.17  &  1.2076 \\
% THEMIS2X_W1.00 & 0.20 &  0.17  &   1.17  &  1.2077 \\

\hline
\end{tabular}
\label{table:tau_J_adhoc}
\end{table}

\begin{figure}
\includegraphics[width=8.5cm]{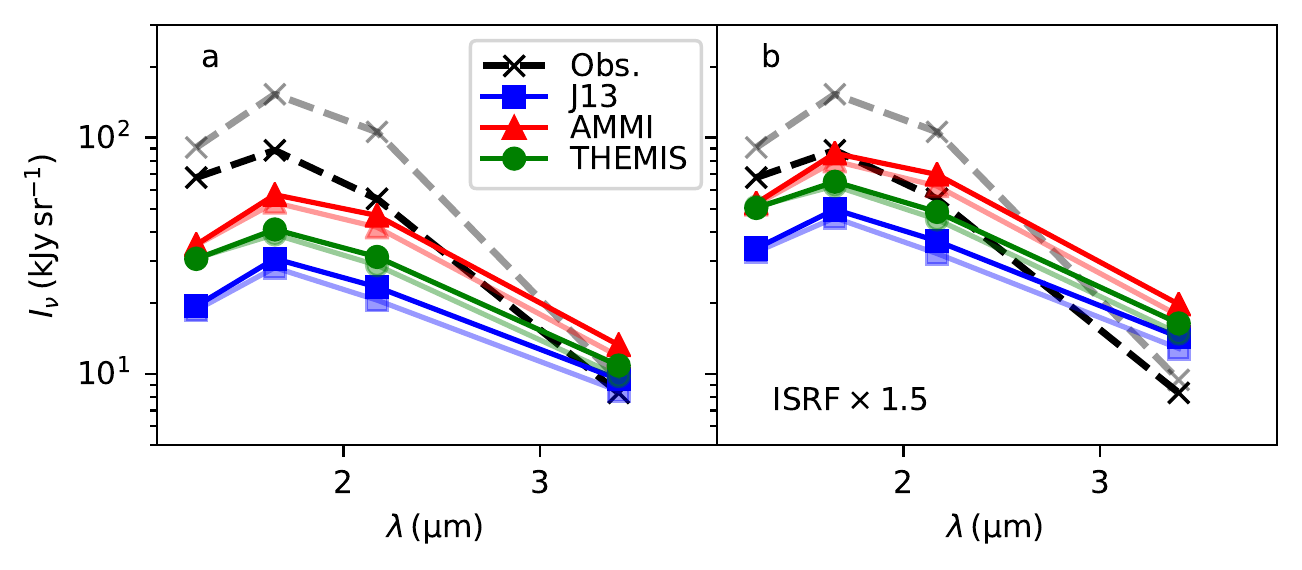}
%
%\sidecaption
\caption{
Comparison of observed net surface brightness (dashed lines) and model
predictions for dust models with ad hoc increase in the sub-millimetre
emissivity. In frame b, ISRF has been scaled with a factor of 1.5.
}
\label{fig:plot_JHK_vs_alternatives_3}
\end{figure}

Figure~\ref{fig:plot_JHK_vs_alternatives_3} shows the NIR surface brightness
predictions for these ad hoc dust models. 
Fig.~\ref{fig:plot_JHK_vs_alternatives_3}a can be compared to
Fig.~\ref{fig:plot_JHK_vs_alternatives_2}a, where, considering the NIR data, the
only difference is in the NIR optical depth (with the factors in the fourth
column of Table~\ref{table:tau_J_adhoc}). The decreased NIR optical depth
provided by the ad hoc dust models is sufficient to bring the NIR signal close to
the observed level for the AMMI dust. The further reduction of the optical depth
by a factor of $\sim$2 in Fig.~\ref{fig:plot_JHK_vs_alternatives_2}a does not
lead to significant additional improvement, apart from the higher values in the
$J$ band. The J13 dust model is still excluded, however, even after the
modifications. It would match NIR observations only with a much stronger
radiation field, which would be in contradiction with the sub-millimetre emission
modelling. For the THEMIS model, the short-wavelength intensities have increased,
bringing them close to the observations.
Comparisons with Fig.~\ref{fig:plot_JHK_vs_alternatives} and
Fig.~\ref{fig:plot_JHK_vs_alternatives_2} show, however, that the increased
brightness cannot be explained simply by the average cloud optical depth. 
Instead, it is partly caused by a change in the column density structure that
has increased the NIR surface brightness contrast relative to the reference area.
In the THEMIS case, the results depend on the densities at which the dust
properties are assumed to change. If the density thresholds were lower by a
factor of two, the contrast between the cloud centre and the reference area would
increase, and the THEMIS spectrum would rise above the AMMI curve.

The modified THEMIS model had $\tau(J)$ close to the observed value
(Table~\ref{table:tau_J_adhoc}) and, together with the pure AMMI case, was
closest to the observed NIR spectrum (Fig.~\ref{fig:plot_JHK_vs_alternatives_3}).
With further tuning of the relative abundances of the dust components, the NIR
signal might be brought to an even better agreement with the observations.  The
modified THEMIS model was made explicitly consistent with the measured NIR vs.
sub-millimetre optical depth ratio, $\tau(250\,\mu{\rm m})/\tau(J)\sim 10^{-3}$.
The high values of $\tau(250\,\mu{\rm m})/\tau(J)$ (shown by direct observations
and required by the modelling) also are qualitatively consistent with recent
laboratory work. In these studies, interstellar dust analogues are shown to have
much higher dust opacities at longer wavelengths ($\lambda>20\,\mu$m) compared to
the current silicate dust models \citep{Demyk2017a}. The laboratory measurements
also reveal a significant temperature-dependence in the sub-millimetre opacities
and opacity spectral indices that should be taken into account in future cloud
models.

\subsection{Comparison to other studies}

\citet{2012A&A...544A..14J} modelled the light scattering and dust emission of
the northern filament of the Corona Australis cloud. Similar to the present
study, the models fitting observations of sub-millimetre dust emission predicted
NIR cloud opacities that were clearly higher than the direct extinction
measurements with background stars. The observed level of NIR scattering could be
matched only by assuming a significant increase in the intensity of the NIR
radiation field. In that paper, models were calculated for two dust models, the
$R_V=5.5$ dust from \citet{Draine2003} and the \citet{OH94} dust with thin
ice mantles. The latter was found to provide a better description of the Corona
Australis filament, which contains a couple of dense clumps with column densities
higher than in LDN~1642. 

\citet{Ysard2013A} modelled dust emission of the Taurus L1506 cloud, including
density-dependent dust evolution. The strong sub-millimetre emission suggested
the presence of dust aggregates at densities above 1500\,cm$^{-3}$.
\citet{Ysard2016} carried out corresponding modelling of dust scattering using
the THEMIS evolutionary dust model, comparing the results to NIR observations and
the MIR Spitzer IRAC data on 21 starless cores. The coreshine observations
required the presence of evolved dust, such as a combination of CMM and AMM
(aggregates without ice mantles) or a combination of CMM and AMMI (aggregates
with ice mantles), the cloudshine data being more compatible with the latter. The
intensity and the balance between NIR and MIR brightness could be adjusted by
changing the density thresholds for the transition between different dust
populations. However, in that study the scattering was not modelled
simultaneously with dust emission, relying on generic spherical density
distributions instead. One of the main conclusions was that, thanks to the H-rich
carbon mantles, the NIR-MIR scattering could be explained with a smaller increase
in the grain volumes. In other earlier studies, coreshine was associated with
grains larger than 1\,$\mu$m \citep{Andersen2013, Steinacker2015}, but also was
seen to be directly linked with the appearance of ice features
\citep{Andersen2014}.

\citet{Togi2017} studied dust emission and NIR-MIR scattering in the cloud B207.
The cloud hosts a single protostar and has a peak column density of $N_{\rm
H2}\sim 3.5\times 10^{22}$\,cm$^{-2}$, which is three times higher than in
LDN~1642. The analysis pointed to high dust albedo values that peak at $A=0.84$
in the $I$ band. The comparison with \citet{Ysard2016} models showed that
observations could be explained best with dust properties similar to CMM+AMM or
CMM+AMMI.  While the models matched the NIR-MIR signal in the cloud core, they
underestimated the sub-millimetre emission by a factor of two. The comparison is
complicated because the peak column density is 2.5 times higher in B207 than in
the model that it was compared with. If the model column density were scaled
upwards, the sub-millimetre surface brightness would not increase proportionally,
because of the simultaneous drop in dust temperatures. Furthermore, as seen in
the LDN~1642 modelling, the predicted NIR intensities would be reduced, because
of the reduction in the number of scattered photons and because of the larger
negative contribution of the $I_{\rm bg}(e^{-\tau}-1)$ term. Therefore, also the
B207 data seem to point towards the dust having a high opacity ratio between the
sub-millimetre and NIR wavelengths.  

In the present paper, we examined signs of dust evolution mostly by comparing the
results for a diffuse-medium dust model, the model J13, and for a dense-medium
dust model, the AMMI model for aggregates with ice mantles. In any realistic
scenario, the dust properties should vary inside the cloud in a continuous
fashion. Our NIR data covered only the central part of the LDN~1642 cloud and
thus do not trace the full transition from diffuse medium to cloud cores. With
optical data over a more extended area, \citet{Mattila2018} estimated in LDN~1642
an $i$-band albedo of $A\sim0.72$ and showed those observations to be consistent
with pure CMM dust. \citet{Saajasto2020} studied the thermal emission and NIR
scattering in the cloud LDN\,1521, also attempting self-consistent modelling of
both FIR emission and NIR scattering. The best fitting models included two or
three dust components and the dust evolution was modelled by modifying their
relative abundances as a function of density. Compared to our results, the
$\tau(\rm J)^{\rm M}\, / \, \tau(\rm J)^{\rm N}$ ratios (NIR optical depths in
emission models vs. direct NIR extinction measurements) reported by Saajasto et
al. were closer to unity, 1.56 and 0.74 in tests with the SIGMA
\citep{Lefevre2019} and THEMIS dust models, respectively. The SIGMA model clearly
overestimated the NIR surface brightness, while the THEMIS model predicted better
the surface brightness in the dense parts of the cloud.

\section{Conclusions} \label{sect:conclusions}

We have examined dust emission, scattering, and extinction in the high-latitude,
star-forming molecular cloud LDN~1642. The new HAWK-I data provided estimates for
the NIR extinction and net surface brightness, which is the sum of scattered
light and attenuated LOS sky background. Together with the Herschel
sub-millimetre maps, these data provided a good starting point for the
testing of different dust models. The study led to the following conclusions:

\begin{itemize}
\item The maximum extinction in LDN~1642 is 
$A_J=2.5$, which corresponds to $A_V=9.3$ ($R_{\rm
V}=3.1)$, at a resolution of 2$\arcmin$. 
\item 
There are no indications of NIR extinction-curve variations; the NIR
colour excesses increase linearly with $N({\rm H}_2)$ up to the
highest column densities, and the observed ratio $E({\rm H-K})/E({\rm
J-H})=0.73\pm0.35$ is consistent with the standard extinction curve. 
\item
We find an optical depth ratio of $\tau(250\,\mu{\rm m})/\tau({\rm
J})\approx 10^{-3}$. This result is similar to previous ratios found for cold
clumps and a few times higher than in the diffuse medium, thus confirming
the increase of the dust sub-millimetre emissivity. 
\item
The sub-millimetre observations could be fitted well with radiative
transfer models, irrespective of the assumed dust model. However,
these result in tens of percent differences in the absolute $N({\rm
H}_2)$ values and the relative values between regions of low and high
column density. 
\item
Compared to the diffuse-medium dust model J13, the evolved dust model AMMI
results in up to 2\,K lower temperatures. This difference is a combined effect of
changes in the sub-millimetre vs. optical opacity ratios and changes
in the absolute opacity values.
\item
The models fitting the sub-millimetre dust emission predict NIR
extinctions that are 2.3-4.6 times higher than the direct extinction
measurements. The discrepancy affects all of the tested dust models
and, in the modelling of the NIR surface brightness, results in SEDs
with too low intensities and wrong spectral shapes.
\item 
With dust properties appropriate for the diffuse medium (dust model J13),
the modelled intensity of the scattered light was at least a factor of three
below that of the observations. This difference remained true even if one assumed a 50\%
higher radiation field and a 50\% lower sky brightness behind the
cloud. This excludes J13 as a viable dust model for LDN~1642.
\item We tested ad hoc variations of the dust models where the $\tau(250\,\mu{\rm
m})/\tau(J)$ ratio was increased to the empirically found value.
The modified J13 dust model was still excluded because of weak NIR
scattering. The modified AMMI and THEMIS models resulted in NIR-MIR
signal almost at the observed level, with approximate agreement also
in the ISRF scaling between the NIR and sub-millimetre models
(Fig.~\ref{fig:plot_JHK_vs_alternatives_3}b).
\item The study shows that LDN~1642 contains evolved dust with 
high sub-millimetre opacity and NIR scattering cross section. The
direct observations of the $\tau(250\,\mu{\rm m})/\tau(J)\approx
10^{-3}$ ratio and the modelling of dust emission and scattering show
that the ratio of dust sub-millimetre and NIR dust opacities is higher
than in the current dust models.
\end{itemize}

In this paper, we have examined dust properties using data on NIR
extinction, NIR scattering, and sub-millimetre emission. Further
crucial and complementary pieces of information may be provided by
future observations, such as James Webb Space Observatory
\citep{Gardner2006} measurements of the MIR ice and silicate features,
or observations with the planned SPICA satellite \citep{Roelfsema2018}
of dust polarisation and MIR-to-FIR dust spectra, clarifying  the
picture of both the large-grain properties and the populations of very
small grains.

\begin{acknowledgements}
NS acknowledges the financial support from the visitor and mobility
program of the Finnish Centre for Astronomy with ESO (FINCA), funded
by the Academy of Finland grant number 306531.
EM is funded by the University of Helsinki doctoral school in
particle physics and universe sciences (PAPU).
VMP acknowledges support by the Spanish MINECO under project AYA2017-88754-P.
\end{acknowledgements}

\bibliography{L1642.bib}

\begin{appendix}

\section{Comparison with DSS data} \label{app:DSS}

\begin{figure*}
\sidecaption
\includegraphics[width=12.0cm]{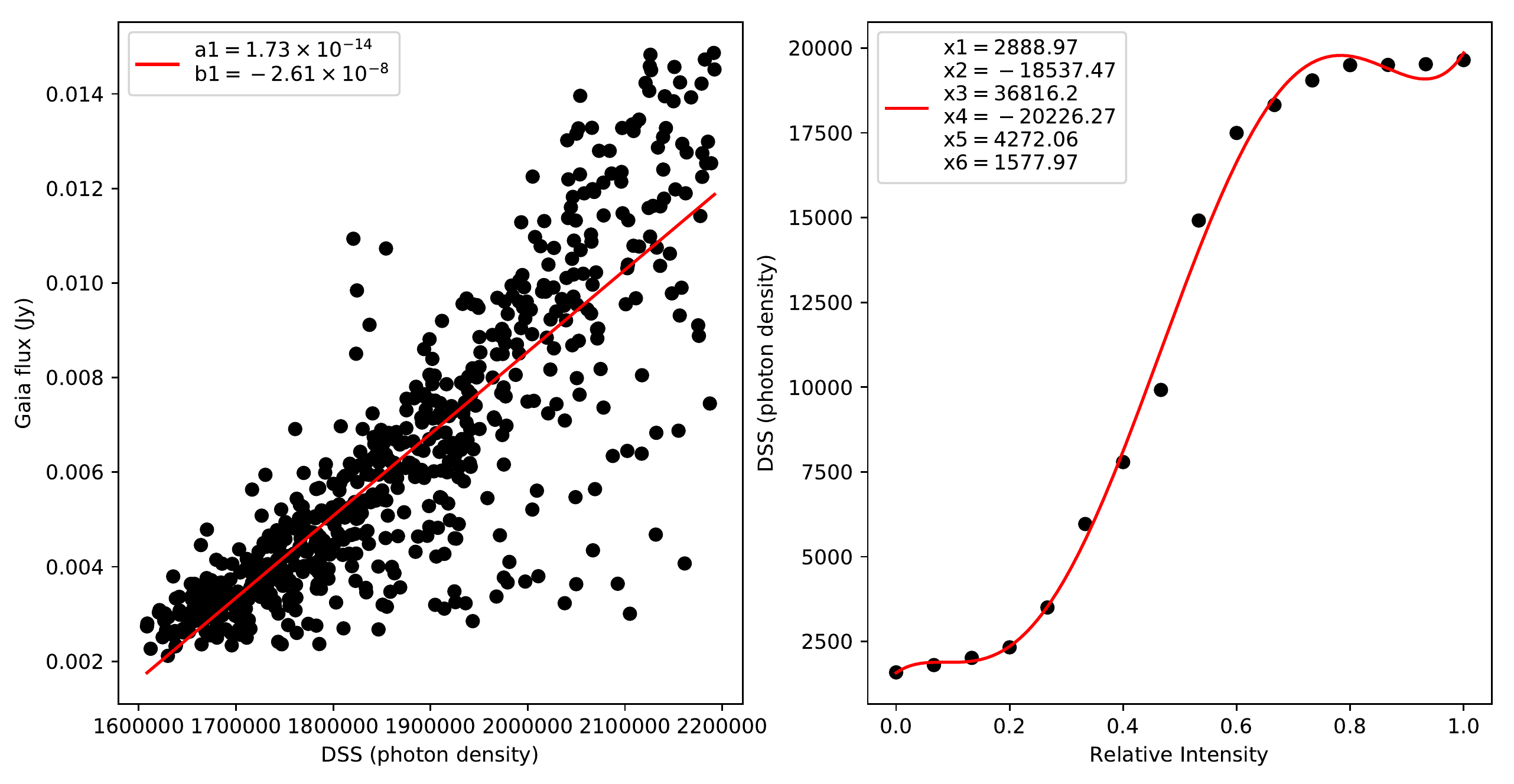} 
\caption{
Fits to the DSS and \textit{Gaia} observations. Left panel: A linear
fit to the relation between the DSS and \textit{Gaia} observations and
the best fit parameters of the fit. Right panel: The response of the
photographic plate from the sensitometer spots of the photographic
plate. The red curve is a fifth order polynomial fit to the
sensitometric data, with the coefficients listed in the figure.
} \label{fig:Gaia_DSS}
\end{figure*}

The digitised DSS images are in units of photon density. We use the
\textit{Gaia} observations to calibrate them to units of Jy sr$^{-1}$,
using photometry from $\sim 1000$ stars from the DR2 catalogue. The
selected stars are within $1.5^\circ$ of LDN~1642, but because the
\textit{Gaia} observations are considerably deeper than the DSS images
and because of the non-linear response of the photographic plates (see
right panel of Fig.~\ref{fig:Gaia_DSS}), we restrict the analysis to
stars with intensities in the range 6500-46000\,$\rm e \; s^{-1}$ in
the red part of the G band filter ($\rm G_r$). We convert the
\textit{Gaia} fluxes to physical units following the \textit{Gaia}
documentation \citep[][chapter 5.3.6]{Gaia_document_5}, by first
converting the flux to an AB magnitude \citep{Oke1983}. The
\textit{Gaia} instrumental magnitude is defined as
%\begin{equation}
$\rm G = -2.5 \log \textit{I} + G_{0,AB}$,
%\end{equation}
where $I$ is the weighted mean flux of the source and $\rm G_{0,AB}$
is the zero point in the AB system. The AB system can be generalised
\citep{Bessell2012} to be used with broad photometric bands so that
%
%\begin{equation}
$\rm AB = -2.5 \log \langle f_\nu \rangle - 56.10$,
%\end{equation}
where $\langle f_\nu \rangle$ is the source mean flux per frequency and the constant
56.10 takes into account the fact that \textit{Gaia} fluxes are in units of
$\rm W \, m^{-2} \, Hz^{-1}$. Combining the two equations we have
%
%\begin{equation}
$<f_\nu> = I \times 10^{-0.4 \, (\rm G_{0,AB} \, + \, 56.10)}$.
%\end{equation}
We use a value of 25.1161 for the $\rm G_{0,AB}$, which is the zero
point of the $\rm G_r$ band \citep{Evans2018}. 
We estimate the DSS photon density flux of the stars using aperture
photometry with a fixed aperture size of 7.5$\arcsec$. The conversion
factors are estimated with a linear fit to the relation between the
\textit{Gaia} fluxes and the DSS photon densities, as shown in the
left panel of Fig.~\ref{fig:Gaia_DSS}a. The right panel in
Fig.~\ref{fig:Gaia_DSS} shows the sensitivity of the DSS photographic
plate, computed as averages over the sensitometric spots on the plate.
%% with the best fit
%% parameters $\rm a1 = 1.73 \times 10^{-14}$ and $\rm b1 = -2.61 \times
%% 10^{-8}$. 
The conversion factors are then used to convert the DSS image to
Jy\,sr$^{-1}$. We assume an uncertainty of $\pm 20 \, \%$ in the
regions where the photon density is in the range 4000-17500\,$\rm e \;
s^{-1}$. The conversion becomes uncertain outside of this range.

\begin{figure}
\includegraphics[width=8.8cm]{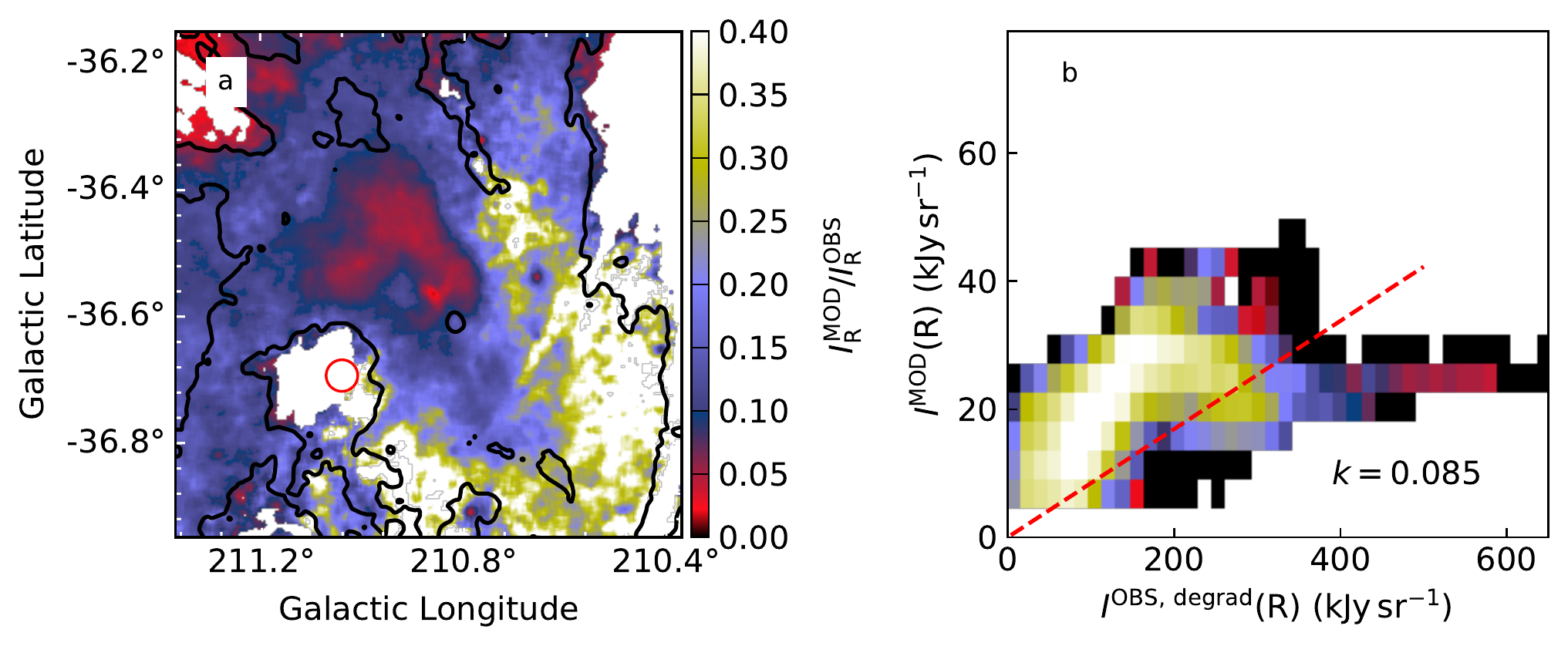}
\caption{
Comparison of observed and modelled (J13, $W$=1) $R$-band surface
brightness. 
Frame a shows the map of the ratio $I_{\nu}^{\rm MOD}/I_{\nu}^{\rm
OBS}$ with contours at $I_{\nu}^{\rm OBS}$ equal to 20\,kJy\,sr$^{-1}$
and 40\,kJy\,sr$^{-1}$ (cf. Fig.~\ref{fig:plot_scattered_light}a). The
red circle indicates the reference area used for establishing a common
zero point.
Frame b shows the correlation as 2d histogram, with a logarithmic
colour scale for the point density. $I_{\nu}^{\rm OBS, decorr}$ stands
for observations corrected for the main gradient. The dashed line
shows the least-squares fit to data with $I_{\nu}^{\rm OBS,
decorr}<300$\,kJy\,sr$^{-1}$. The effect of background sky brightness
is not included in the model.
}
\label{fig:cmp_DSS_model}
\end{figure}

We computed $R$-band predictions only for scattered light because we do
not have estimates for the absolute sky brightness $I_{\rm bg}$ at
this wavelength. Figure~\ref{fig:cmp_DSS_model} shows that the RT
model predictions are one fourth of the observed sky brightness. The
ratio $I_{\nu}^{\rm MOD}/I_{\nu}^{\rm OBS}$ shows a gradient, which
could be an artefact from the DSS plate (Fig.~\ref{fig:DSS_area})
Figure~\ref{fig:cmp_DSS_model}b shows that, after removing the mean
gradient (some 1.5\% per arcmin), the least-squares slope is even
lower, $\Delta I_{\nu}^{\rm MOD}/\Delta I_{\nu}^{\rm OBS} \approx
0.09$. It may be biased towards lower values by residual contribution
from point sources (high values of observed intensity). On the other
hand, any surface brightness in the background sky would decrease the
slope further. Therefore, the default models definitely fail to
produce enough surface brightness in the R band.

\section{Further model calculations}  \label{app:sca}

\subsection{Dust emission models}  \label{app:emission}

\begin{figure*}
\includegraphics[width=18cm]{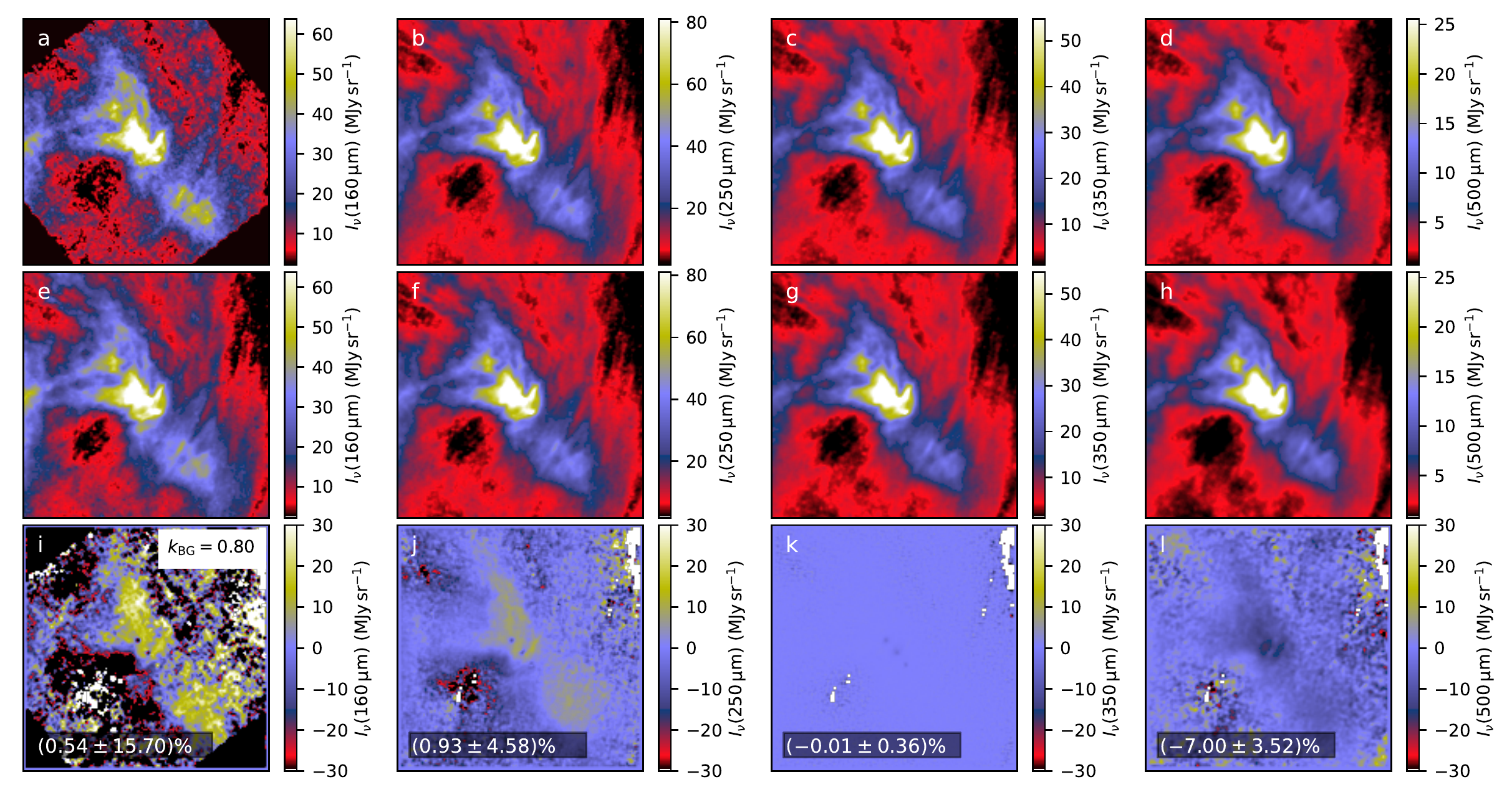}
%
%\sidecaption
\caption{
Sub-millimetre surface brightness in L1642. The observed 250\,$\mu$m, 350\,$\mu$m, and
500\,$\mu$m maps (frames a-d) and the surface brightness predicted by
the models with J13 dust with $W=1$ (frames e-h). Frames i-l show the
relative errors of the fits.
}
\label{fig:mod_emit}
\end{figure*}

Figure~\ref{fig:mod_emit} compares the sub-millimetre observations to
the modelling with J13 dust. The RT models were optimised to match the
250-500\,$\mu$m data but the figure also shows a comparison with the
160\,$\mu$m data. 

\begin{figure*}
\includegraphics[width=18cm]{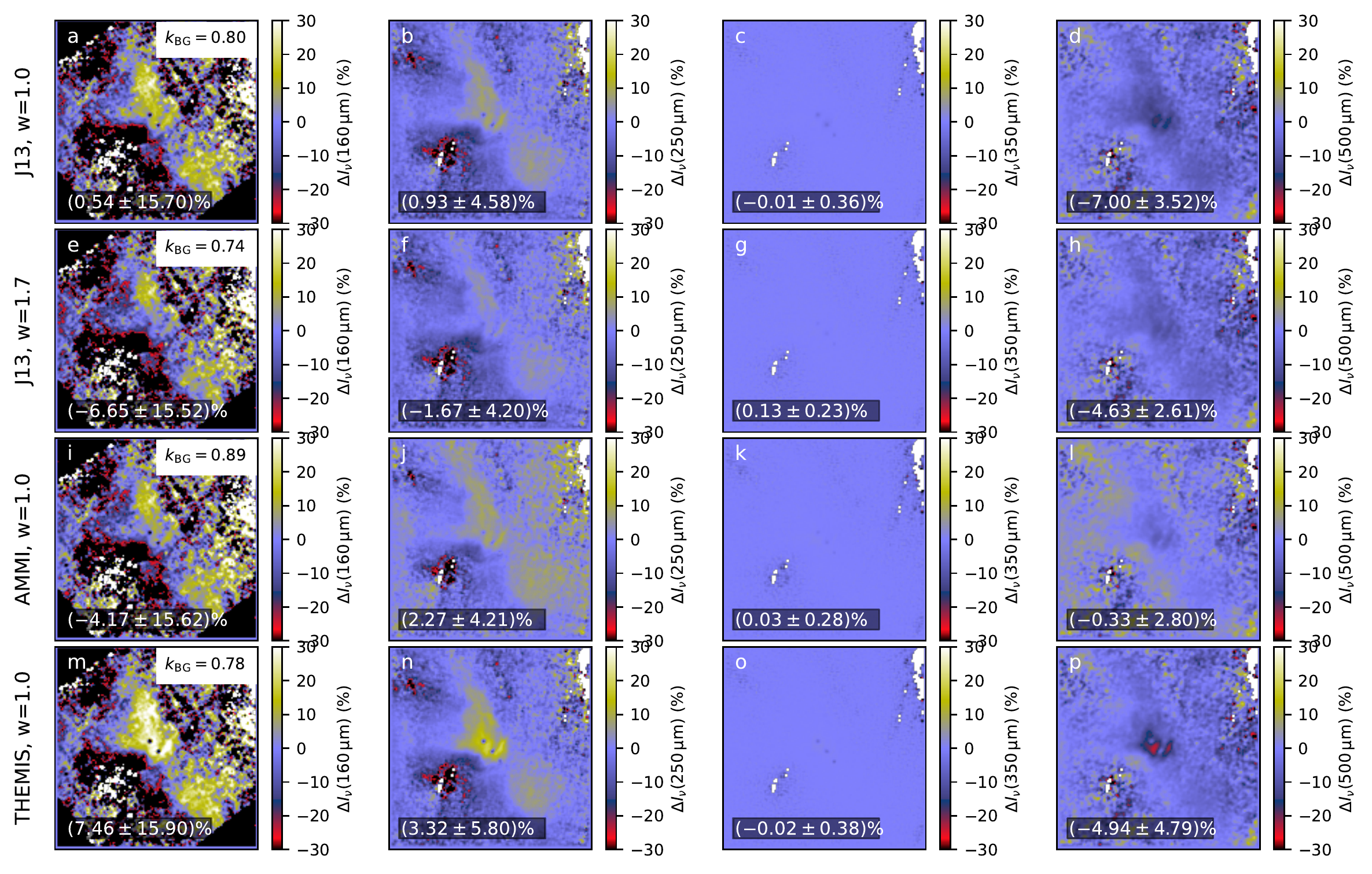}
%
%\sidecaption
\caption{
Residuals of the modelled 250-500\,$\mu$m surface brightness,
$(I_{\nu}^{\rm OBS}-I_{\nu}^{\rm MOD})/I_{\nu}^{\rm OBS}$. The columns
correspond to 160\,$\mu$m, 250\,$\mu$m, 350\,$\mu$m, and 500\,$\mu$m, respectively. The
160\,$\mu$m data were not used in the model optimisation. The rows
correspond to models with different dust (J13, AMMI, and THEMIS) and
LOS cloud extent ($W=1.0$ or $W=1.7$). The average value and the standard
deviation of the relative residual is quoted at the bottom of each frame.
}
\label{fig:mod_emit_2}
\end{figure*}

Figure~\ref{fig:mod_emit_2} shows the fit residuals for alternative
models with different LOS cloud extents ($W$=1.0 and 1.7) and dust
models J13, AMMI, and THEMIS. Except for the THEMIS model, the dust
properties are constant throughout the model volume. The fits are
found to be of similar quality, although with more variation in the
160\,$\mu$m residuals.

\subsection{Predictions of extended scattering}

\begin{figure*}
\includegraphics[width=18.0cm]{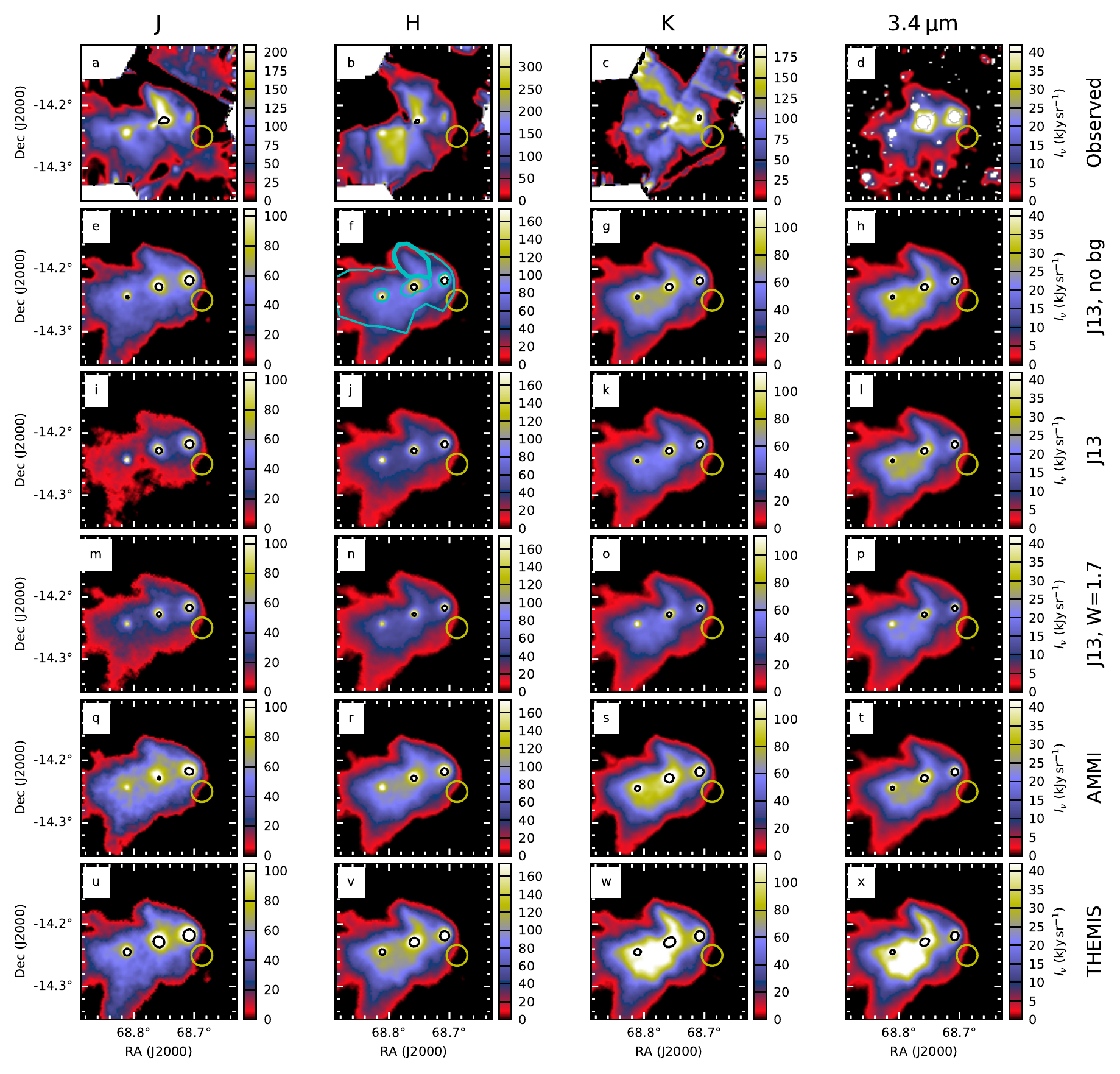}
\caption{
Comparison of 1.25-3.4\,$\mu$m surface brightness between observations
and models. The figure is the same as Fig.~\ref{fig:cmp_JHK_model_all}
but the model clouds have lower column densities that correspond to
the NICER NIR extinction measurements. The colour scales are the same
for all model plots of the same band. 
%% The black contours are at 2, 3, 4, and 8 times the maximum value of the colour scale.
The yellow circles indicate the reference region used for background
subtraction. The black contours are drawn at 1.5 times the maximum of the
colour scale.
}
\label{fig:cmp_JHK_model_all_rescaled}
\end{figure*}

Figure~\ref{fig:cmp_JHK_model_all} showed predictions of NIR surface
brightness for the ISRF intensity of \citet{Mathis1983}. 
Figure~\ref{fig:cmp_JHK_model_all_rescaled} shows the same comparison
when the cloud optical depths have been scaled down to match the
average NICER extinction.

\subsection{Scattering near embedded sources}

\begin{figure}
\includegraphics[width=8.8cm]{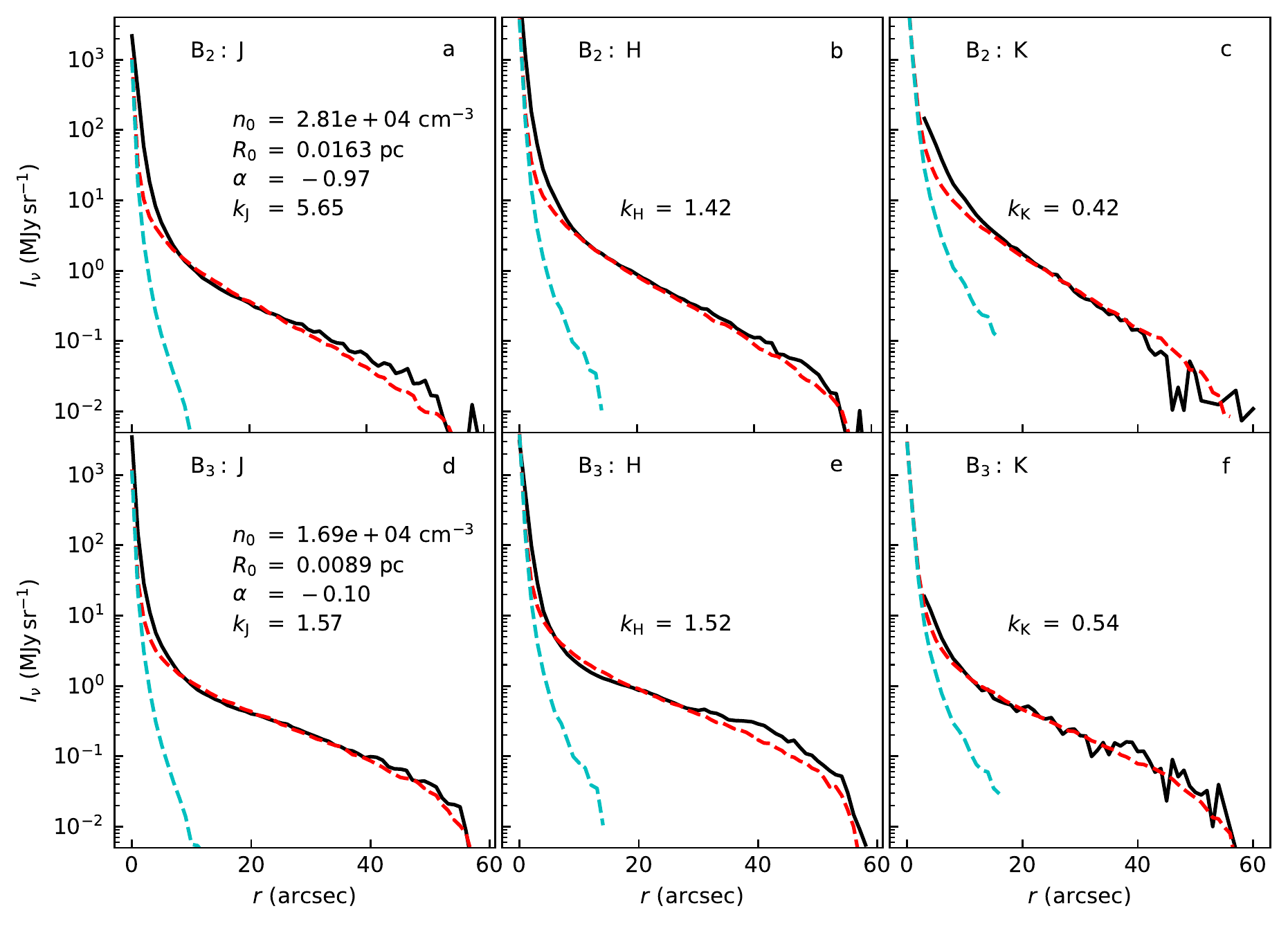}
%
%\sidecaption
\caption{
Surface brightness profiles in the vicinity of the embedded sources B2 (frames
a-c) and B3 (frames d-f), for the $J$, $H$, and $K_S$ bands.  This plot is
the same as Fig.~\ref{fig:plot_sphere} but with the AMMI dust model
\citep{Ysard2016}.
}
\label{fig:plot_sphere_AMMI}
\end{figure}

Section~\ref{sect:PS_models} showed results for scattered light from
spherically symmetric models of the source B2 and B3 environment,
based on the use of the J13 dust model \citep{Compiegne2011}.
Figure~\ref{fig:plot_sphere_AMMI} show the corresponding results for
the AMMI dust models.

\end{appendix}

\end{document}